\documentclass{pasj00}
\usepackage{multirow}
\usepackage[varg]{txfonts}

\begin{document}
\SetRunningHead{Mitsuishi et al.}{
X-ray Emission Associated with the Shapley Supercluster}
\Received{2011/05/31}
\Accepted{2011/09/14}

\title{
Search for X-Ray Emission Associated with the Shapley Supercluster
with {\it Suzaku}
}

\author{Ikuyuki  \textsc{Mitsuishi}\altaffilmark{1} , Anjali Gupta\altaffilmark{2}, 
Noriko Y. Yamasaki\altaffilmark{1}, 
Yoh Takei\altaffilmark{1}, Takaya Ohashi\altaffilmark{3}, \\Kosuke Sato\altaffilmark{4}, 
 Massimiliano Galeazzi\altaffilmark{2},
J. Patrick Henry\altaffilmark{5}, and Richard L. Kelley\altaffilmark{6}
}
\affil{\altaffilmark{1}Institute of Space and Astronautical Science, Japan Aerospace Exploration Agency (ISAS/JAXA),\\
3-1-1 Yoshinodai, Chuo-ku
Sagamihara, Kanagawa, 252-5210}
\email{mitsuisi@astro.isas.jaxa.jp}

\affil{\altaffilmark{2}Physics Department, University of Miami,
1320 Campo Sano Drive Coral Gables, \\FL 33146, USA}

\affil{\altaffilmark{3}Department of Physics, Tokyo Metropolitan University, 
1-1 Minami-Osawa, Hachioji, \\
Tokyo, 192-0397}

\affil{\altaffilmark{4}Department of Physics, Tokyo University of Science, 
1-3 Kagurazaka, Shinjyuku-ku, Tokyo 162-8601}

\affil{\altaffilmark{5}Institute for Astronomy, University of Hawaii, 2680 Woodlawn Drive, Honolulu, HI 96822, USA}

\affil{\altaffilmark{6}NASA/Goddard Space Flight Center, Greenbelt, MD 20771, USA}
%


%

\KeyWords{X-ray, cluster, supercluster, WHIM} 

\maketitle

\begin{abstract}
{\it Suzaku} performed observations of 3 regions in and around the Shapley supercluster: a region located between A3558 and
A3556, at $\sim 0.9$ times the virial radii of both clusters,
and two other regions at $1^\circ$ and $4^\circ$ away from the
first pointing.  The $4^\circ$-offset observation was used to evaluate
the Galactic foreground emission. 
We did not detect significant redshifted Oxygen emission lines 
(O \emissiontype{VII} and O \emissiontype{VIII}) in the spectra of
all three pointings, after subtracting the contribution
of foreground and background emission.  An upper limit for the
redshifted O \emissiontype{VIII} K$\alpha$ line intensity
of the warm-hot intergalactic medium (WHIM) is $1.5 \times 10^{-7}$ photons s$^{-1}$ cm$^{-2}$
arcmin$^{-2}$, which corresponds to an overdensity of $\sim 380\
(Z/0.1~Z_{\odot})^{-1/2} (L/3~{\rm{Mpc}})^{-1/2}$, assuming
$T=3\times10^6$ K.
We found excess continuum emission in the $1^\circ$-offset and
on-filament regions, represented by thermal models with $ kT\sim 1$
keV and $\sim 2$ keV, respectively.  The redshifts of both 0 and that
of the supercluster (0.048) are consistent with the observed spectra.
The $\sim 1$ keV emission can be also fitted with Ne-rich Galactic
(zero redshift) thin thermal emission.
Radial intensity profile of 2~keV component suggests contribution from
A3558 and A3556, but with significant steepening of the intensity
slope in the outer region of A3558.
Finally, we summarized the previous {\it Suzaku} search for the 
WHIM and discussed the
feasibility of constraining the WHIM\@.
An overdensity of $< 400$ can be detectable using O\emissiontype{VII}
and O\emissiontype{VIII} emission lines in a range of $1.4\times
10^6~\mathrm{K}<T<5\times 10^{6}~\mathrm{K}$ or a continuum emission
in a relatively high temperature range $T>5\times 10^{6}~\mathrm{K}$
with the {\it Suzaku} XIS.  The non detection with {\it Suzaku}
suggests that typical line-of-sight average overdensity is $< 400$.
\end{abstract}

\section{Introduction}
\label{SEC:introduction}
History of the Universe has been revealed little by little but constantly. 
Some fundamental mysteries are, however, remaining.
One of these unresolved questions is the
so-called ``missing baryon" problem. 
More than half of the total baryonic matter 
in the current epoch ({\it z}$<$1) has not been detected, and is
known as the ``missing baryons".
Cosmological hydrodynamic simulations (e.g.,
\cite{cen-ostriker,dave})
predict that most of missing baryons are expected to be
in the phase of warm-hot intergalactic medium (WHIM),
a plasma existing along filamentary large-scale structures connecting 
clusters of galaxies.
The WHIM may be detected in UV and X-rays,
since its expected temperature is 10$^5$ to 10$^7$ K.
However, due to its low density and low temperature on the order of
10$^{-5}$ to 10$^{-3}$ cm$^{-3}$ and 10$^5$ to 10$^7$ K, technically, it
is difficult to achieve a sufficient
sensitivity,  especially in the X-ray energy.

Many observational progresses on the WHIM has been recently reported. 
In the UV band, O\emissiontype{VI} absorption lines ($\lambda=1032, \ 1038
~\mathrm{\AA}$) have been detected along the lines of sight to $\sim 50$
AGN (e.g.,
\cite{2005ApJ...624..555D,2008ApJ...679..194D,2008ApJS..177...39T}),
by {\it FUSE} and Hubble Space Telescope.
If we assume all the observed O\emissiontype{VI} lines are produced by the WHIM,
then the O\emissiontype{VI} can probe 7--10~\% of the baryons
\citep{2005ApJ...624..555D,2006ASPC..348..341T,2009AIPC.1135....8D}.
With the inclusion of the thermally broadened Ly$\alpha$ absorbers
(BLAs) recently observed by \citet{2010ApJ...710..613D} and
\citet{2010ApJ...712.1443N,2010ApJ...721..960N}, these estimates are
boosted up by a factor of $\sim$2, i.e., UV observations are capable of
probing $\sim$20\% of the baryons in the local universe
\citep{2010ApJ...710..613D}.
In X-ray band, there are many observations to investigate the absorption
lines from highly ionized ions, but they lead to only several marginal
($<4\sigma$) detection of absorption lines associated with dense parts
of large-scale structures
(\cite{2004PASJ...56L..29F,2007ApJ...655..831T,2009ApJ...695.1351B,
2010ApJ...714.1715F}).  Two examples of the possible WHIM signal
are an excess X-ray continuum (modeled as $T=$0.9~keV plasma) emission in a
filamentary region between A222 and A223 by XMM-Newton (\cite{werner}) 
and a signature on the angular correlation of diffuse X-ray emission 
in six accumulated intergalactic filaments (\cite{galeazzi1}).
Emission lines of O\emissiontype{VII} and O\emissiontype{VIII} in X-ray
spectra can also be used as a signature of the WHIM.  However, it is
challenging for present instruments because of the low intensity of the
thin plasma and because of the difficulty of distinguishing it from the
soft Galactic emission.  Nevertheless, thanks to the good sensitivity
and line spread function at the lower energy below 1 keV of the XIS (X-ray Imaging Spectrometer;
\cite{koyama-xis}) detectors on board {\it Suzaku} (\cite{mitsuda-Suzaku}), the good constrains on the WHIM
properties has been obtained (\cite{intensity-overdensity,
virial3}).  {\it Suzaku} XIS is also used as a powerful tool to
extend the ICM study to outskirt regions around virial radii
(\cite{bautz-a1795,george-pks0745-191,virial1,virial2,Simionescu-2011}).


In this paper, we report on observations in a supercluster as a
target to search for the WHIM emission.  Superclusters are 
thought to be promising targets because they
are known as the most largest and densest structures associated with the
large scale structures inducing a WHIM concentration and because
the faint WHIM
emission is accumulated along the line-of-sight depth of the order of
several Mpc or more.
Among them, we focused on the Shapley supercluster ({\it z} = 0.048) using {\it Suzaku}.

Throughout this paper, we adopt H$_0$ = 73 km s$^{-1}$ Mpc$^{-1}$,
$\Omega_M$ = 0.73 and $\Omega_\Lambda$ = 0.27, which corresponds to 57
kpc arcmin$^{-1}$ at the redshift of 0.048.
Unless otherwise specified, all errors are at 90 $\%$ confidence level.
   

\section{Observations \& Data Reduction}
\subsection{Observations}
\begin{figure}[htbp]
\begin{center}
\FigureFile(70mm,70mm)
  {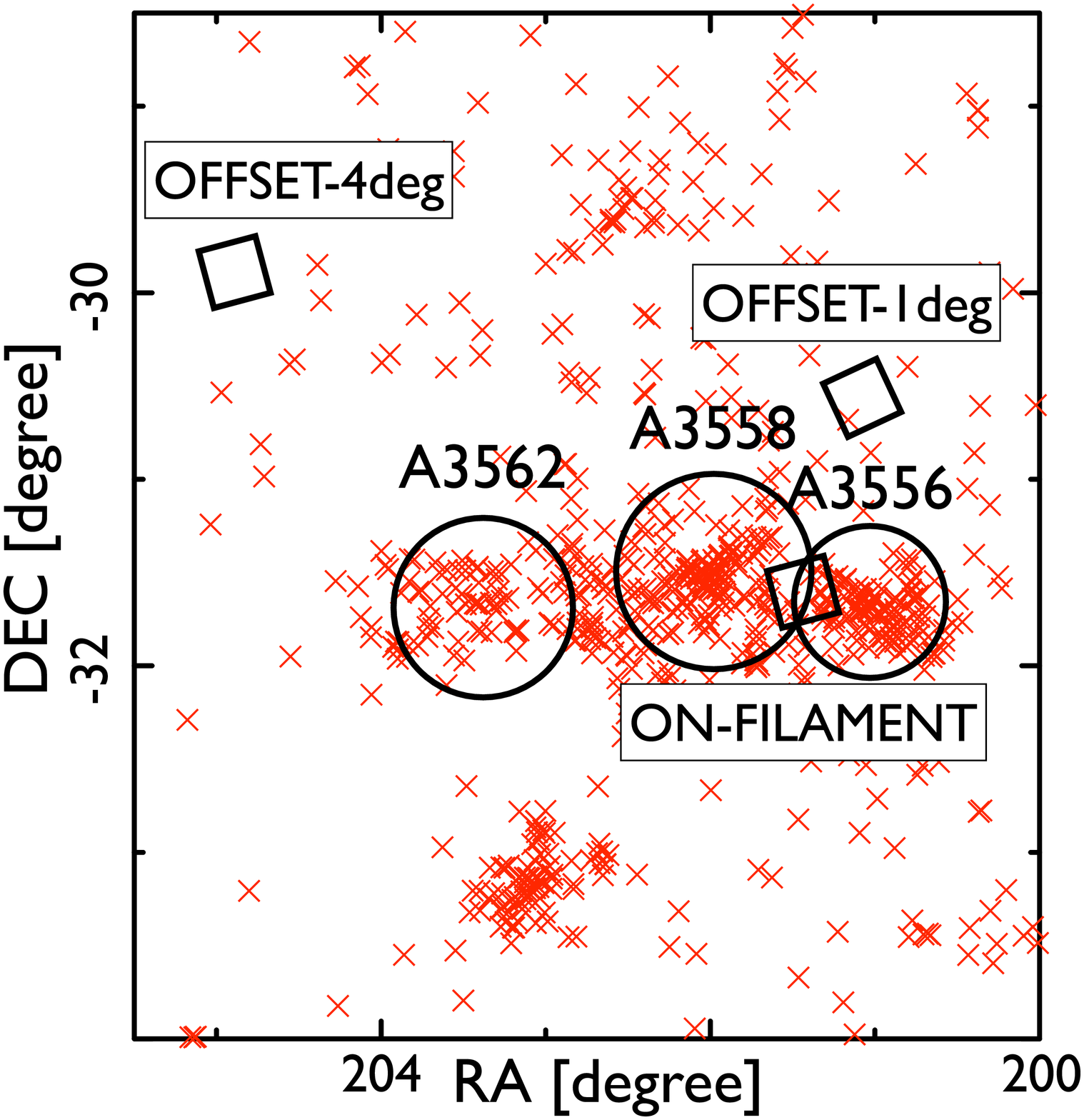}
\end{center}
\caption{The distribution of galaxies with known redshifts
  (red crosses) in the range of $0.046 < z <  0.050$. 
  Circles indicate the virial radius of each cluster. 
  The virial radius of A3558 and A3562 are adopted in \citet{a3558-a3562-virial-radius} and A3556 is 
  calculated using the equation (1) in \citet{virial2}.
  Three squares show the fields of view of {\it Suzaku XIS} in the
  OFFSET-1deg, OFFSET-4deg and ON-FILAMENT pointings.}
\label{fig:observation-positions}
\end{figure}

Three pointings in and around the Shapley supercluster were carried out
with {\it Suzaku} in
July 2008 (PI: Ohashi \& Galeazzi).
The Shapley supercluster is one of promising candidates 
to search for an emission from the WHIM
because this is one of the largest (\cite{shapley1,shapley2})
and most massive (\cite{shapley3}) structures in the local Universe
on the order of several 10 Mpc and $10^{16}\ M_{\odot}$,
respectively.
Actually, ROSAT detected an excess emission between 0.5 and 2 keV in
the Shapley supercluster above the surrounding fields (\cite{rosat}),
which may indicate the emission of WHIM origin.  Thus we observed the
same region between A3556 ($kT\sim 3.2$ keV, \cite{a3558-akimoto}) and
A3558 ($kT\sim 5.6$ keV, \cite{a3558-akimoto}), located close to the
center of the Shapley supercluster (hereafter we call the field as
ON-FILAMENT)\@.
In addition, two offset regions at $\sim 1^\circ$ (hereafter OFFSET-1deg)
and $\sim 4^\circ$ (OFFSET-4deg) from the ON-FILAMENT region
were observed to constrain the foreground
emission around the supercluster.
Observed regions are shown in Figure \ref{fig:observation-positions}
with the distribution of galaxies whose redshifts are known with the
NASA Extragalactic Database (NED) and in the range of $0.046 < z <
0.050$. This redshift range includes the nearby clusters A3556 ($z =
0.0479$), A3558 ($z = 0.048$) and A3562 ($z = 0.049$) in the Shapley
supercluster, and the corresponding velocity width is $\sim 600$ km
s$^{-1}$ at $z = 0.048$.  It is not completely
  surveyed, and the magnitude is ranging from 13.5 to 20.3, 14.2 to
  17.2 in R-band, 14.0 to 15.5 in I-band and 12.6 to 20.4 in V-band
  respectively.
The sequence numbers, observation dates, pointing directions,
exposure times and the hydrogen column densities are 
summarized in Table \ref{table:observation-log}. 
We confirmed that the difference between the Galactic hydrogen column density 
by \citet{lab-survey} from the LAB survey and \citet{Galactic-absorption}
was only less than 10 \% on the order of several 10$^{20}$ cm$^{-2}$, 
which leads to a systematic error of $<$3 \% 
for the best-fit values in the analysis.
This was smaller than the statistical errors.
In the spectral fitting shown in this paper, 
we always fixed the N$_H$ value at the LAB survey values because of its better angular resolution.
\begin{table*}[htbp]
\begin{center}
\small
\caption{Observation log}
\label{table:observation-log}
\begin{tabular*}{16cm}{llcccc} \hline\hline
  &      & ON-FILAMENT  &       OFFSET-1deg     & OFFSET-4deg      \\   \hline
Sequence number  &                             & 803072010        &       803068010 & 803021010   \\  
Obs date   &                                              & 2008-07-10 to 2008-07-13   &  2008-07-19 to 2008-07-23  &2008-07-18 to 2008-07-19 \\
($\alpha_{2000}$,$\delta_{2000}$)  & degree &  (201.5, -31.6) & (201.1, -30.6)  & (204.9, -29.9) \\
 ($\ell$, $b$) & degree &  (311.4, 30.7) &  (311.3, 31.8)   & (315.2, 31.8)   \\
Exposure & ks   & 30.2$^{\ast}$ &   143   &  47.2 \\
$N_{\rm H}$ & cm$^{-2}$  & 4.15 $\times$10$^{20}$ &   4.78 $\times$10$^{20}$ 
&3.98 $\times$10$^{20}$\\
\hline 
\end{tabular*}
\begin{flushleft} 
\footnotesize{\hspace{1cm}$^{\ast}$ After the COR screening as described.}
\end{flushleft}
\end{center}
\end{table*}
\subsection{Data Reduction}
In this work, we analyzed only the XIS data.  The data reduction and
analysis were carried out with HEAsoft version 6.11 and XSPEC 12.7.0 with AtomDB ver.2.0.1.  
We produced the redistribution matrix files (RMFs) by xisrmfgen
ftool, in which a degradation of energy resolution and its position
dependence are included.  We also prepared ancillary response files
(ARFs) using xissimarfgen ftool (\cite{arf}).  An input image to
xissimarfgen is a uniform circle of a radius of $20'$, except for ARFs
for point source candidates.  For the ARFs of the point source
candidates, we simply assumed point source images at the source
positions in the sky.  As a systematic uncertainty in the thickness of the
XIS optical blocking filter, we considered $\pm 20\%$.  With these uncertainties,
we examined the significance of the excess component as described in
\S~3.  To reproduce the non X-ray background (NXB) component, we
utilized an accumulated dark Earth database with the "xisnxbgen"
ftools task (\cite{xisnxbgen}).

In addition to standard data processing (e.g.\ ELV $>5$ and DYE\_ELV
$>20$), we examined if additional filtering was necessary.  
Firstly, we checked a contamination of a fluorescent scattering of 
the neutral oxygen line from the atmosphere of the Earth.
The flux of the line is characterized by a total column density of neutral oxygen atoms 
and molecules in a sun-lit atmosphere (e.g., \citet{swcx-atomcol}).
We adopted a column density threshold of 10$^{14}$ cm$^{-2}$ and compared the spectra 
including and excluding the duration in which the column density of sun-lit atmosphere is 
above the threshold.
The two spectra were consistent with each other within statistical errors.

Secondly, we examined contaminant emission from the solar wind charge exchange (SWCX) near the Earth, 
which is caused by ions in the Solar wind penetrating into the geomagnetic field and atmospheric neutral hydrogen 
(\cite{swcx1, swcx2,yoshino-san}).
We applied the same screening procedure as described in \cite{hagihara-san}.
The parameters we considered were a proton flux from the sun and an Earth-to-magnetopause 
distance in the line of sight of {\it Suzaku}.
The higher proton flux indicates higher ion flux that produces a contamination of specific lines 
through the CX process as describe in \citet{swcx-protonflux}.
We take 4$\times$10$^8$ cm$^{-2}$ s$^{-1}$ as a threshold of proton flux.
During times with the smaller Earth-to-magnetopause distance, more larger charged particles 
are able to reach the vicinity of the Earth, which results in the larger contamination \citep{swcx1}.
A threshold of five times the Earth radius away from the Earth center is adopted.
Then we compared the screened spectra with the non-screened one.
The two spectra turned out to be consistent with each other within the statistical errors.
Therefore, we did not remove these durations but utilized all the observed data.

Thirdly, we performed a data screening with the cut-off-rigidity (COR) of the Earth's magnetic field 
for the ON-FILAMENT observation, because we found discrepancy between FI and BI sensors 
in the ON-FILAMENT spectra in 2.0--5.0 keV, which may be due to the fluctuation of the detector background.
The COR value corresponds to the threshold momentum preventing a charged particle 
from reaching the surface of the Earth.
A particle (e.g., cosmic rays) whose momentum is smaller than the COR value is screened 
by the Earth's magnetic field.
When COR is small, more particles with smaller momentum can reach to the detectors, 
which leads to a higher particle background and larger systematic errors \citep{xisnxbgen}. 
The screening was done based on COR2 parameter \citep{xisnxbgen}.
The spectra from different sensors became consistent with each other within the statistical errors 
if the selection of COR2 $>$ 12 GV was applied.
For OFFSET-1deg and OFFSET-4deg regions, we did not impose this screening 
since no discrepancy was found between the results with and without the COR screening. 

After the above screenings, we removed point source candidates 
in each region with a radius of $1'.2$ to $2'$ depending on the source
intensity.
We picked up point sources in the XIS1 image in 0.4--2.0~keV, based on
wavelet algorithm (by a CIAO tool wavdetect) for point-like sources.
We also carried out visual inspection and extracted somewhat
diffuse-like sources which were not effectively detected with this
algorithm.  As a result, the numbers of point source candidates found
are 5, 8 and 7 for the ON-FILAMENT, OFFSET-1deg and OFFSET-4deg
regions, respectively.

\begin{figure*}
\begin{minipage}{0.325\hsize}
\begin{center}
\FigureFile(63mm,63mm)
  {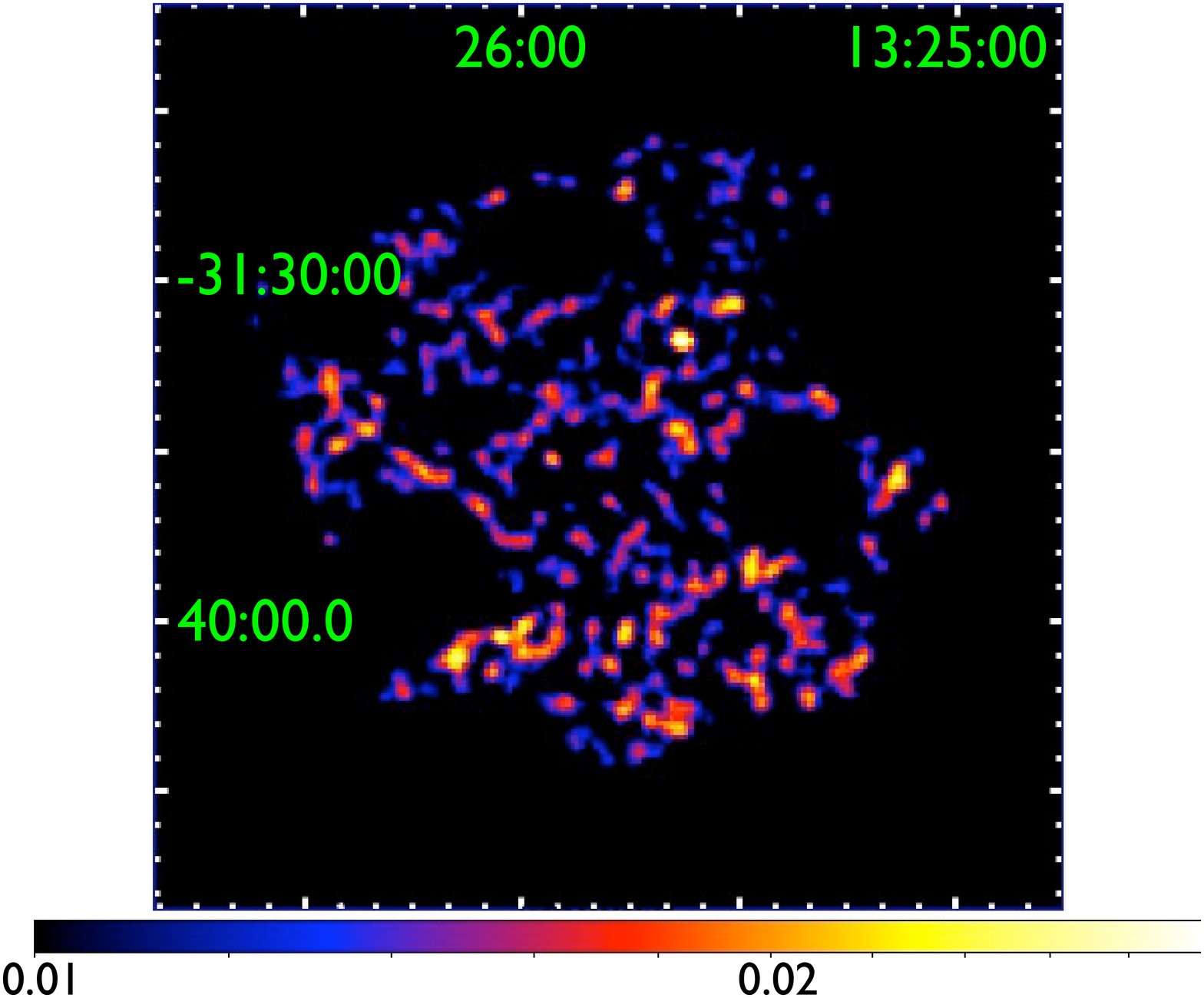}
   \end{center}
   \end{minipage}  
   \begin{minipage}{0.325\hsize}
\begin{center}
 \FigureFile(63mm,63mm)
  {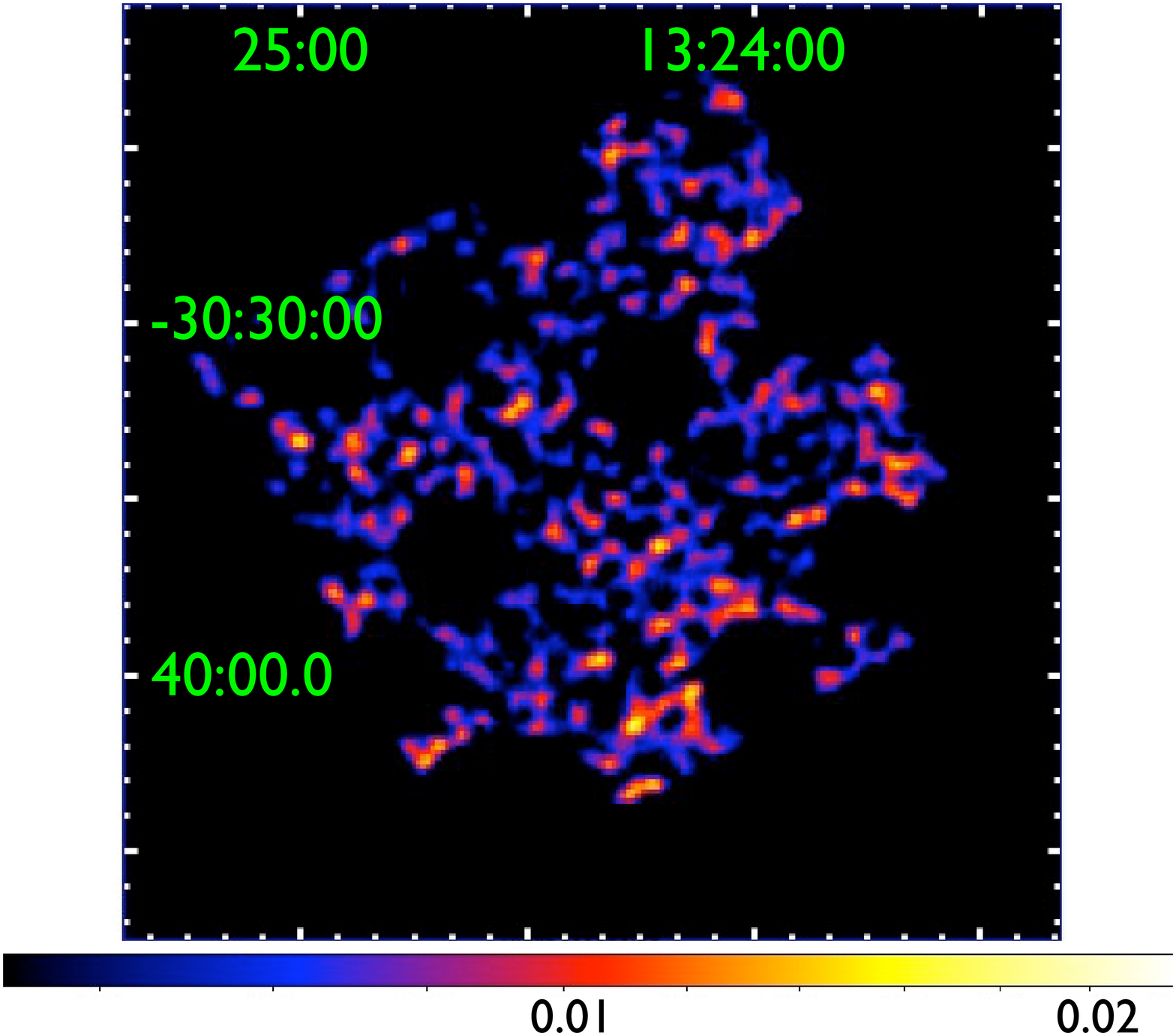}
\end{center}
 \end{minipage}
  \begin{minipage}{0.325\hsize}
\begin{center}
  \FigureFile(63mm,63mm)
  {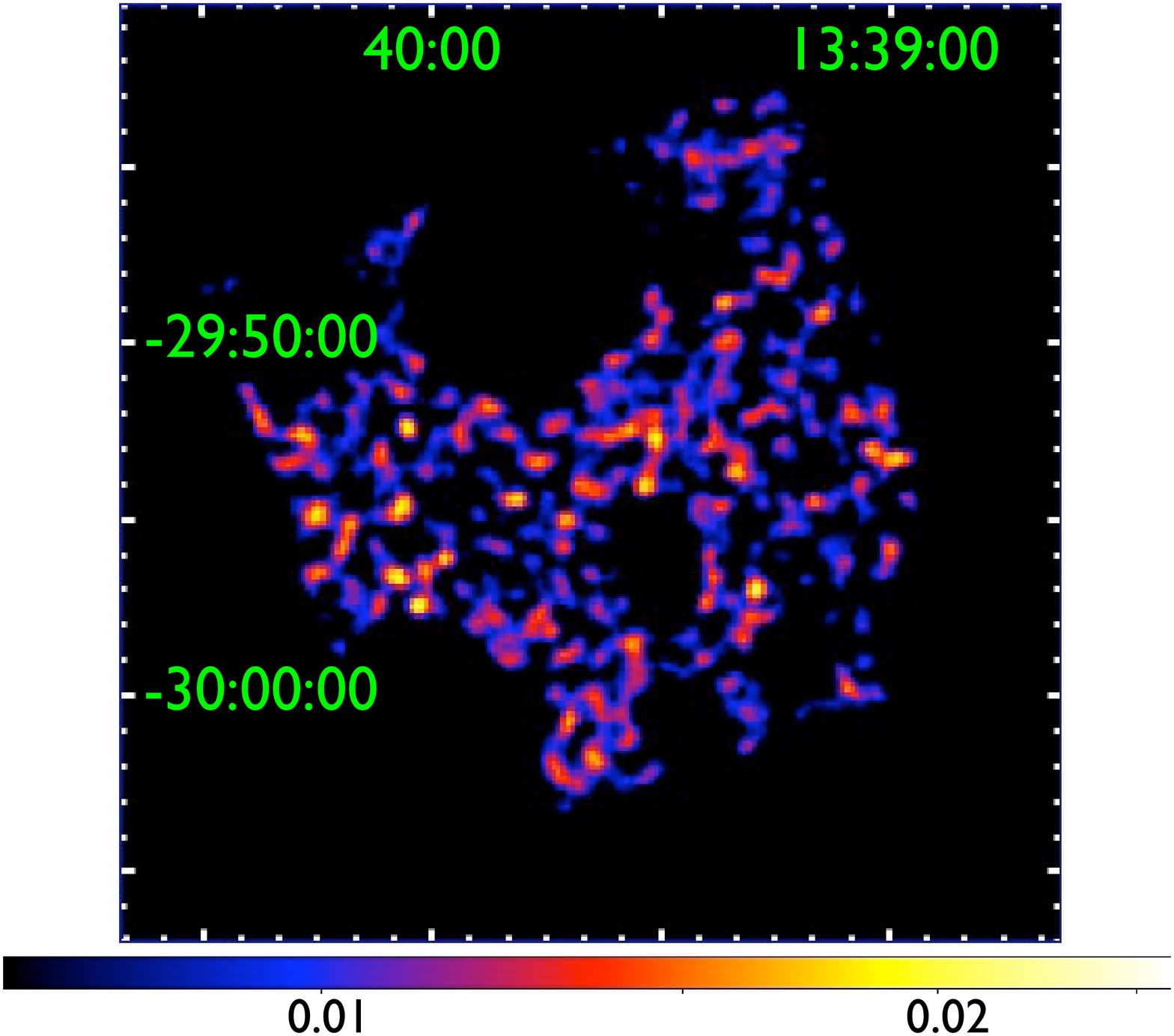}
 \end{center}
   \end{minipage} 
   \caption{The XIS1 images for the ON-FILAMENT (left), OFFSET-1deg
     (middle) and OFFSET-4deg (right) regions. The energy range is
     0.4--2.0~keV in units of cts ks$^{-1}$ (64 pixel)$^{-1}$ .  
     The images are smoothed with a kernel of
     $\sigma=8.3$ arcsec and the scale is logarithmic.  Vignetting and
     background are not corrected for and cal sources are
       removed. The FOV of {\it Suzaku} XIS is $18'\times 18'$.}
\label{fig:xis1-images}
   \end{figure*}

Then, in order to determine the detection limit, we evaluated
NXB-subtracted fluxes of these sources. 
The detection limit will be utilized to estimate the flux and expected fluctuation of 
the CXB (i.e., unresolved point sources) in \S 3.2.
The spectrum of each point source was produced, followed by subtraction of
the NXB spectrum at the same detector region. The NXB spectrum was
re-normalized by adjusting the count rate in 10--15~keV to the source
intensity. 
The re-normalization is applied to the whole spectrum. Note
that the effective area of XRT in 10--15~keV is so small that all the
counts in the source data can be considered as the NXB contribution.
The re-normalization factor is less than 5\% in most cases.
To improve the spectral fit, we adopted this re-normalization to the
NXB spectrum. However, the best-fit parameters did not
change significantly even without this process.
Next, we fitted the NXB-subtracted spectra of point source candidates
in 2.0--10.0~keV with an absorbed power-law model.  
The source fluxes in 2.0--10.0~keV are larger than 
1.2 $\times$10$^{-13}$ ergs cm$^{-2}$ s$^{-1}$ for the ON-FILAMENT, 
2.8 $\times$10$^{-14}$ ergs cm$^{-2}$ s$^{-1}$ for the OFFSET-1deg and 
4.1 $\times$10$^{-14}$ ergs cm$^{-2}$ s$^{-1}$ for the OFFSET-4deg.
XIS1 images in 0.4--2.0~keV from which the regions of point source
candidates are excluded are shown in Figure \ref{fig:xis1-images}.
We removed cal sources by status filter in XSELECT located
  in the edge of a CCD detector in order to analyze the data in the
  energy band above 5 keV\@. We fitted the spectra of the FI and BI
sensors simultaneously, but in different energy ranges which were
0.4-8.0 keV for the BI and 0.5-8.0 keV for the FI, respectively, after
the above data reduction.
\section{Analysis \& Results}
\label{SEC:analysis--results}

\subsection{Background/foreground Emission Analysis}

\begin{figure*}
\begin{minipage}{0.5\hsize}
\begin{center}
\FigureFile(80mm,80mm)
  {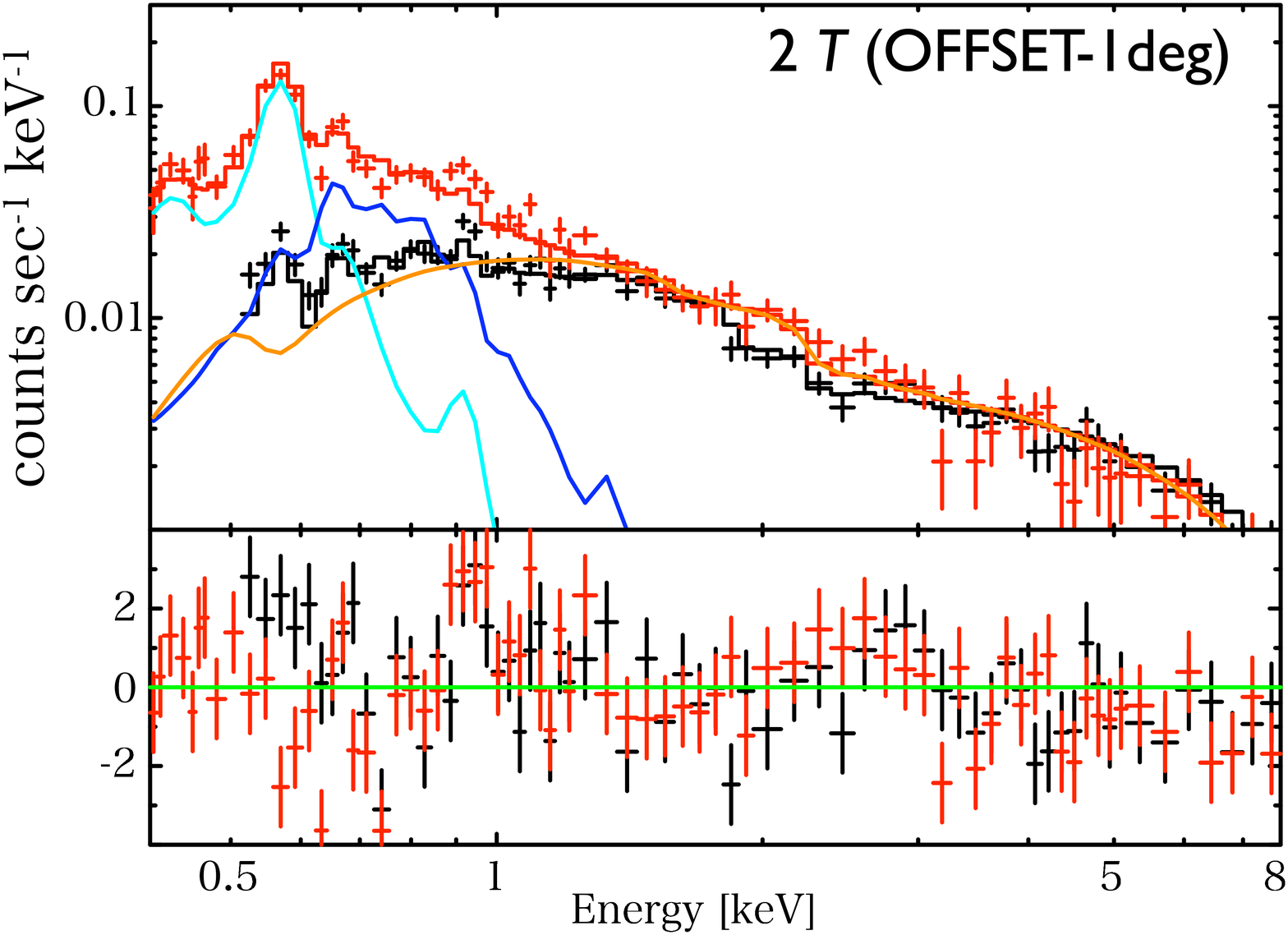}
   \end{center}
   \end{minipage}  
   \begin{minipage}{0.5\hsize}
\begin{center}
 \FigureFile(80mm,80mm)
  {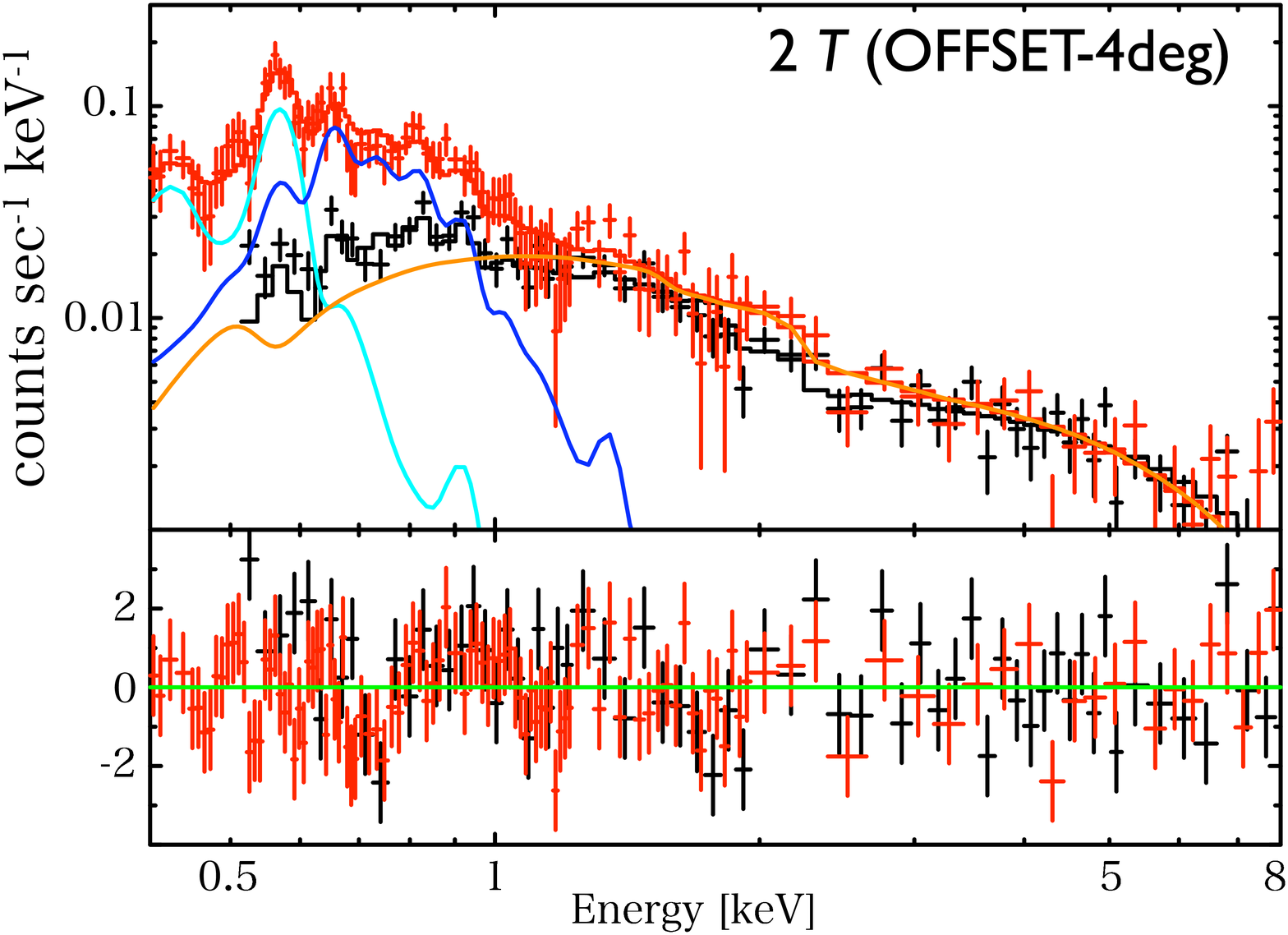}
\end{center}
 \end{minipage}
   \caption{
Spectra in 0.4--8.0~keV
of the OFFSET-1deg (left) and the OFFSET-4deg (right) regions.
Red and black data are for the BI and FI sensors, respectively.
The lines show the best-fit model that consists of
(1) {\it apec$_\mathrm{LHB}$} (cyan), 
(2) {\it phabs}$\times$ {\it apec$_\mathrm{MWH}$} (blue),
and (3) {\it phabs}$\times$ {\it power-law} (orange).
The components (1), (2), and (3) correspond to LHB+SWCX,
MWH and CXB, respectively.
   The photon index of the CXB is fixed to be 1.4.}
   \label{fig:OFFSET-1deg-2-2t}
   \end{figure*}

Since the expected WHIM emission is soft and weak, 
a careful treatment of the background and foreground emission
is mandatory.
Thus firstly, we analyzed the data in the offset regions (OFFSET-1deg
and OFFSET-4deg).
We fitted the spectra with a sum of  (1) an unabsorbed thin thermal
collisionally-ionized equilibrium (CIE)
plasma, (2) an absorbed thin thermal
CIE plasma and (3) an absorbed power law, modeled as \citet{Galactic}.
These first two plasmas represent contributions from
the local emission (SWCX and  Local Hot Bubble: LHB) 
and Galactic halo (MWH: Milky Way Halo), respectively. 
The abundance is set to 1 solar for both models.
The component (3) corresponds to the accumulation of unresolved
extragalactic point sources (cosmic X-ray background: CXB), which is
described by an absorbed power-law model with a photon index of 1.4
in \cite{cxb-kushino}.  
In summary, we used the following models:
{\it apec$_\mathrm{1}$} + {\it phabs}$\times$ 
({\it apec$_\mathrm{2}$}+ {\it power-law}).
These three components are known as the typical
X-ray background and foreground emission.
The best-fit parameters and spectra are shown in Table
\ref{table:OFFSET-1deg-2} (in columns labeled as Nominal)
and in Figure \ref{fig:OFFSET-1deg-2-2t}.
The temperatures of the two thermal plasmas ($\sim 0.1$~keV and $\sim
0.25$~keV in both regions) are typical values reported in
\citet{yoshino-san}, which studied 14 {\it Suzaku} spectra in the blank
sky fields.
\cite{cxb-kushino} discussed the CXB surface brightness using a
$\log N-\log S$ relation.  According to the relation and the present
detection limits in the two regions, the normalizations are expected
to be 6.8 and 7.3 photons s$^{-1}$ cm$^{-2}$ sr$^{-1}$ keV$^{-1}$ at 1
keV, in OFFSET-1deg and OFFSET-4deg regions, respectively. These
values are consistent with the observed ones.

For each region, we evaluated line centers and surface brightnesses of
O\emissiontype{VII} K$\alpha$ and O\emissiontype{VIII} K$\alpha$
emission lines.  In the best-fit model of each region, we replaced the
$\it{apec}$ model with a $\it{vapec}$ model whose Oxygen abundance was
set to be 0 and added two Gaussian lines.  These lines represented
O\emissiontype{VII} and O\emissiontype{VIII} emission lines.  The
abundances of other elements were fixed to be 1 solar, and
temperatures of the vapec components were fixed to the best-fit values
for the individual fits. The line centers and the surface brightnesses
were derived from these spectral fits, and the results are summarized
in Table~\ref{table:ovii-oviii-norm-center}.  The line centers are
consistent with zero redshift within statical errors and within the
typical energy determination accuracy of XIS\@.  The implication for
the line intensity will be discussed in
\S~\ref{SEC:upper-limit-intens}.

\begin{figure*}
\begin{minipage}{0.32\hsize}
\begin{center}
\FigureFile(52mm,52mm)
  {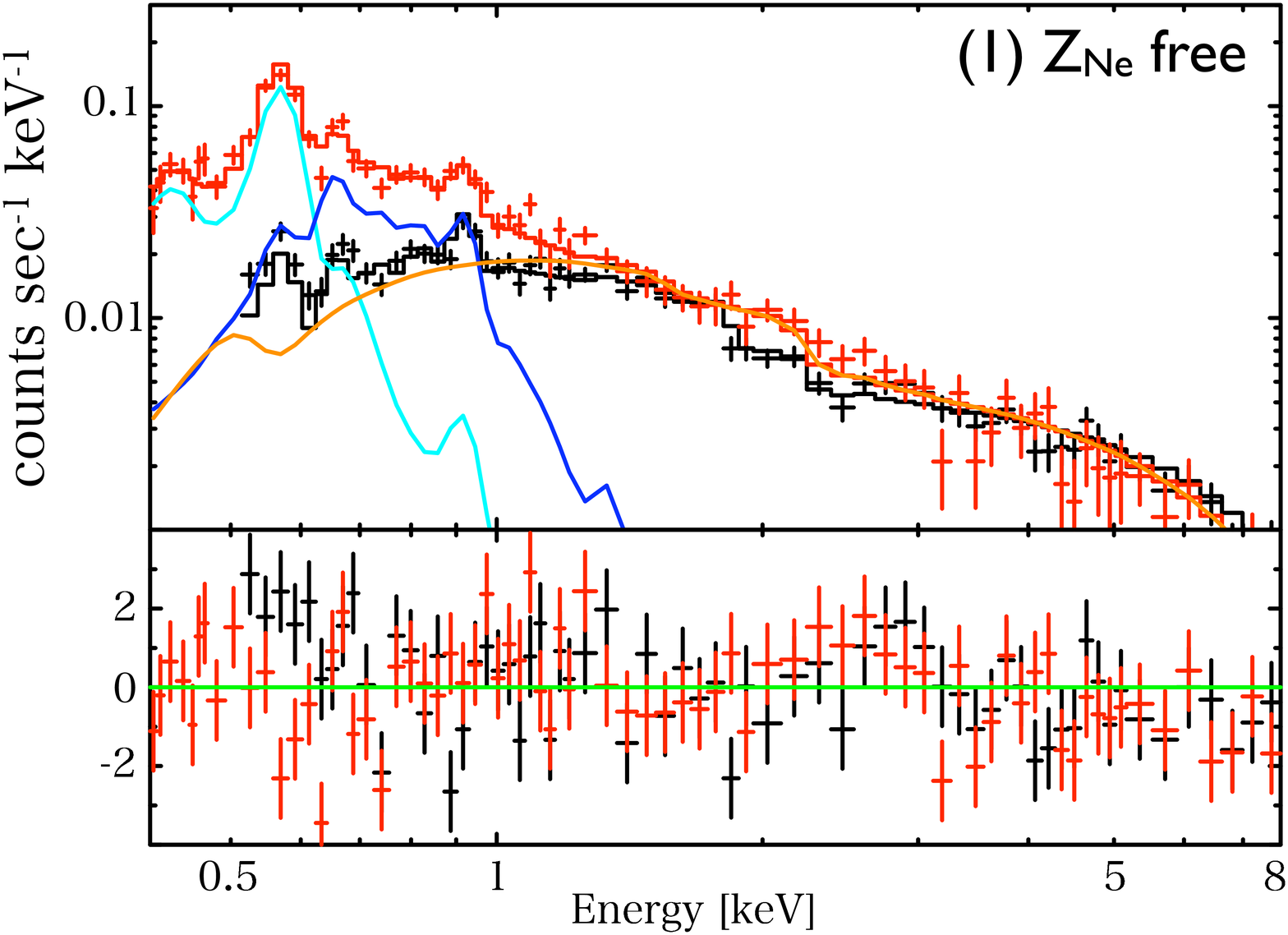}
   \end{center}
   \end{minipage}  
   \begin{minipage}{0.32\hsize}
\begin{center}
 \FigureFile(52mm,52mm)
  {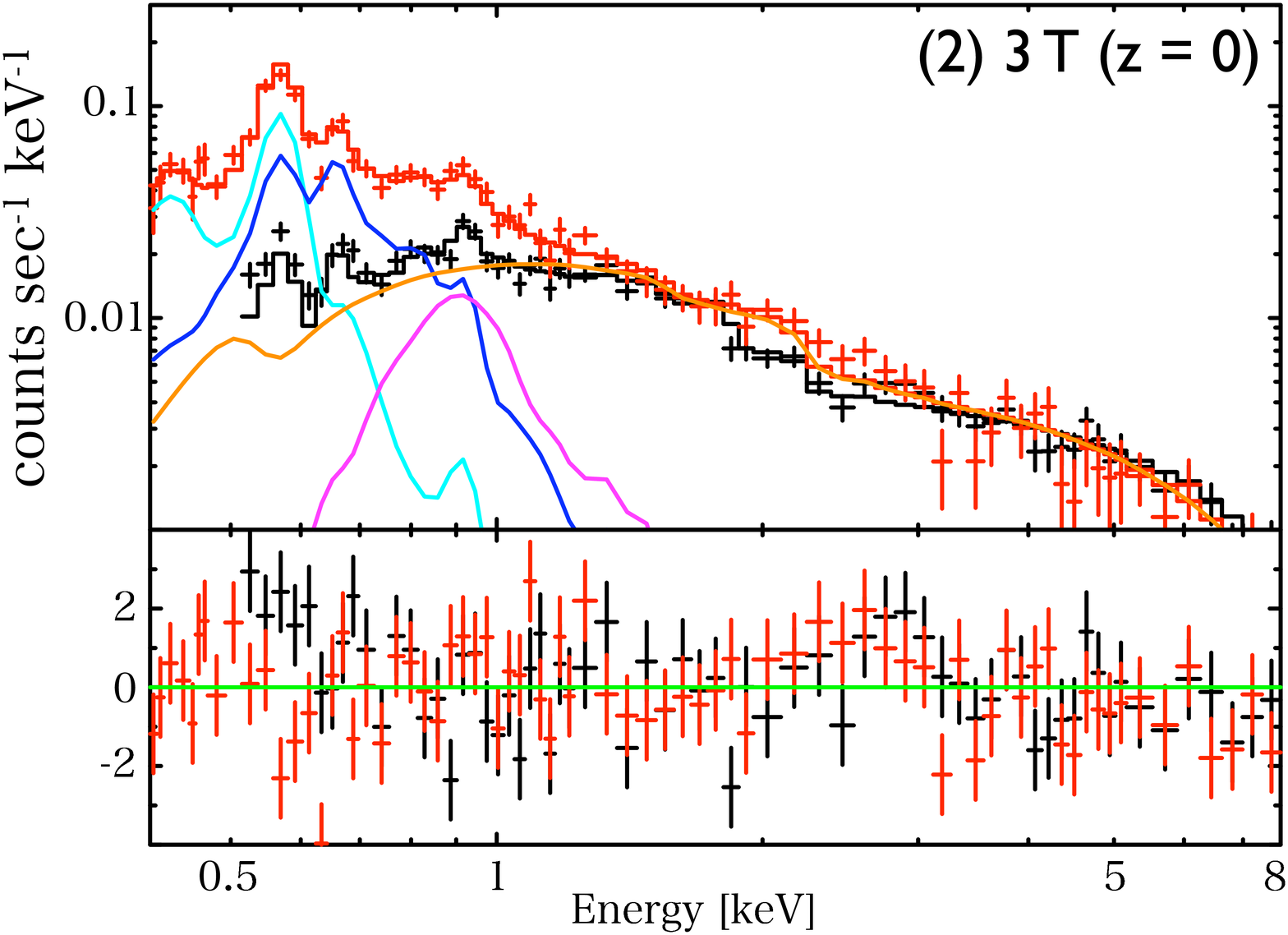}
\end{center}
 \end{minipage}
 \begin{minipage}{0.32\hsize}
\begin{center}
\FigureFile(52mm,52mm)
  {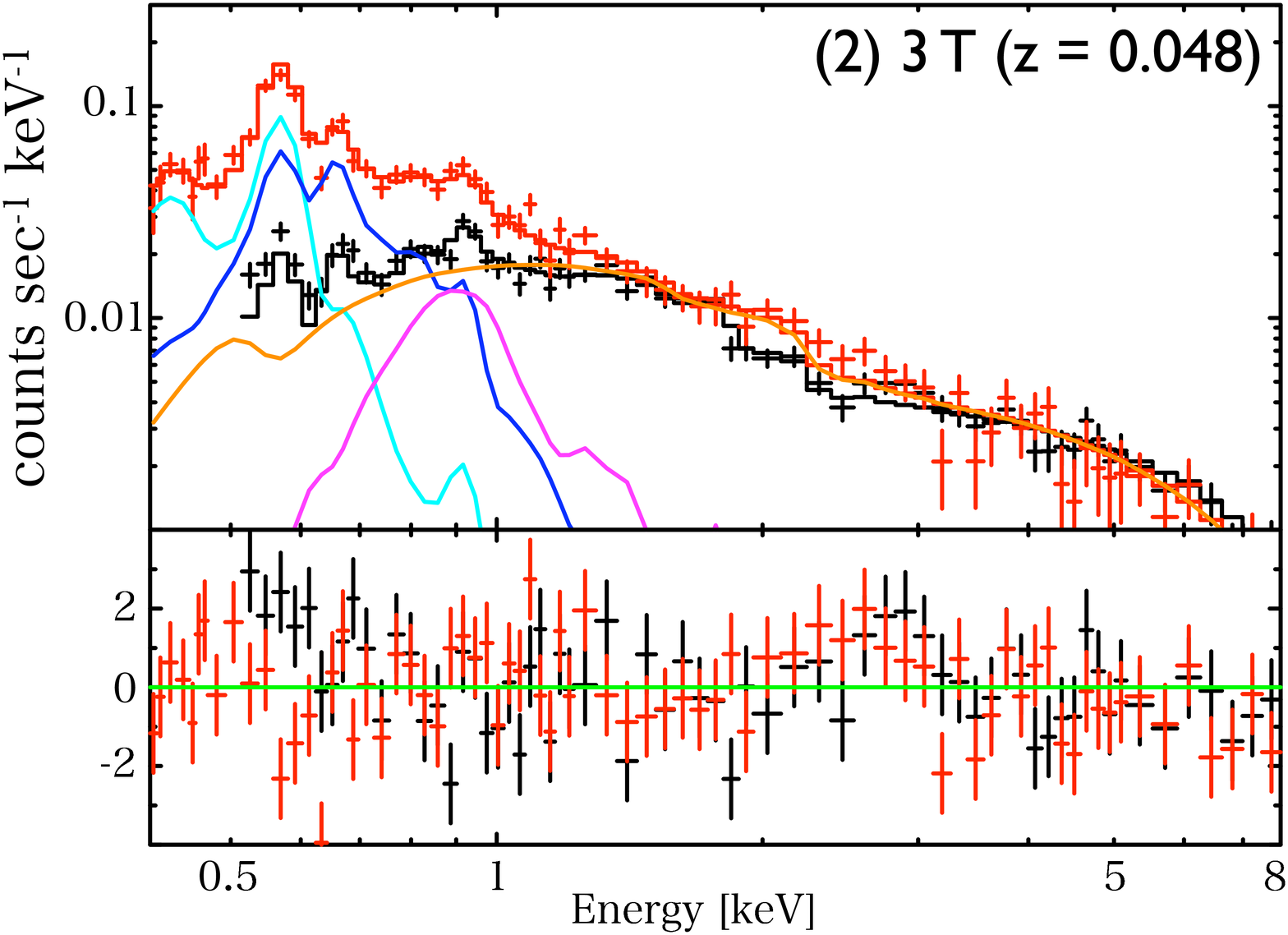}
   \end{center}
   \end{minipage}  
   \caption{Spectral fits for the OFFSET-1deg data in 0.4--8.0~keV
   with different models.
Red and black data are for the BI and FI sensors, respectively.
 (left) (1) + (2) + (3) but Ne
   abundance of the MWH was set to be free to model the residual at around
   0.9 keV\@. (middle) (1) + (2) + (3) + one more higher temperature
   plasma at redshift 0. (right) (1) + (2) + (3) + one more higher
   temperature plasma at the redshift of the supercluster.  
 See caption of Fig.~\ref{fig:OFFSET-1deg-2-2t}
 about the components (1), (2) and (3).
Cyan, blue, orange and magenta lines indiate LHB+SWCX,
   MWH, CXB, and the
   additional higher temperature plasma, respectively.
   The photon index of the CXB is fixed to be 1.4.}
   \label{fig:OFFSET-1deg-2}
   \end{figure*}

We also looked into the possible excess continuum in both the spectra.
While addition of another thermal model did not improve the fit for
OFFSET-4deg spectrum, it slightly improved the fit for the OFFSET-1deg
one.  This is shown in the column $3T\ (z=0)$ of
Table~\ref{table:OFFSET-1deg-2}.  The temperature of the additional
component is $\sim$0.9~keV\@.  A similar improvement was obtained by adding
a thermal emission with the supercluster redshift ($3T\ (z=0.048)$ of
Table~\ref{table:OFFSET-1deg-2}). Since this
excess component is apparently peaked around 0.9 keV, we also tried a
model with different metal abundance. Since Ne lines are concentrated
in this energy, we tried to fit with a vapec model for the MWH
component with variable Ne abundance and other elements fixed at 1
solar done in \citet{yoshino-san}. As shown in the column ($Z_\mathrm{Ne}$ free of
Table~\ref{table:OFFSET-1deg-2}), this model also showed an equally
good fit.  
The spectra fitted with these models are shown in Figure
\ref{fig:OFFSET-1deg-2-2t} and \ref{fig:OFFSET-1deg-2}.
A Ne-rich or higher temperature ($\sim$0.7 keV) Galactic emission 
is reported in some of high-latitude fields, but not the same as the OFFSET-1deg region, 
studied in \citet{yoshino-san}, which shows that we can not deny the possibility of 
the Galactic emission origin.
\renewcommand{\baselinestretch}{1.5}\selectfont
\begin{table*}[htbp]
\begin{center}
\caption{Best fit parameters of the offset regions.}
\label{table:OFFSET-1deg-2}
\begin{tabular}{cccccc} \hline\hline
 
&\multicolumn{4}{c}{OFFSET-1deg} & \multicolumn{1}{c}{OFFSET-4deg} \\
                                                   & Nominal (2 $T$)                      & $Z_\mathrm{Ne}$ free           & 3 $T$ ($z=0$)                          & 3 $T$ ($z=0.048$)                  & Nominal (2 $T$)\\    \hline
$kT_1$ [keV]                           &  0.111$^{+0.009}_{-0.015}$ & 0.103$^{+0.010}_{-0.008}$ & 0.093$^{+0.012}_{-0.008}$   &0.093$^{+0.012}_{-0.009}$   &0.091$\pm$0.014 \\
$\it{Norm_1}$$^{\ast}$          & 56.4$^{+33.2}_{-11.0}$         & 71.2$^{+23.8}_{-18.6}$        & 82.3$^{+24.4}_{-25.5}$          &82.5$^{+39.7}_{-25.1}$          &110.8$^{+120.1}_{-47.5}$ \\
$kT_2$ [keV]                           & 0.250$^{+0.028}_{-0.022}$  & 0.232$^{+0.020}_{-0.012}$ & 0.188$^{+0.021}_{-0.010}$   & 0.185$^{+0.027}_{-0.010}$  &0.240$^{+0.021}_{-0.015}$\\
$Z_\mathrm{Ne}$ (Z$_{\odot}$)& 1.0 (fix)                                & 2.1$\pm$0.3                            & 1.0 (fix)                                      & 1.0 (fix)                                      & 1.0 (fix)\\
$Norm_2$$^{\ast}$                & 7.8$^{+2.5}_{-2.0}$                & 8.6$^{+1.4}_{-2.0}$               & 15.1$^{+6.2}_{-5.2}$               & 15.8$\pm$6.0                          &15.3$\pm$3.0\\  
$\Gamma$ (fix)                       & 1.4 (fix)                                      & 1.4 (fix)                                     &  1.4 (fix)                                      & 1.4 (fix)                                     &1.4 (fix) \\
 $\it{SB}$$^\dagger$             & 6.4$\pm$0.2                             & 6.4$\pm$0.2                            & 6.1$\pm$0.2                             & 6.1$\pm$0.2                            &7.5$\pm$0.3 \\ \hline
$kT_3$                                     & -                                                  & -                                                 & 0.861$^{+0.068}_{-0.091}$   & 0.966$^{+0.099}_{-0.088}$  & -\\
Redshift                                    & -                                                  & -                                                 & 0                                                  & 0.048 (fix)                                & -\\
$Norm_3$$^{\ast}$                & -                                                  & -                                                 & 1.0$^{+0.3}_{-0.2}$                 & 1.5$\pm$0.3                            & -\\ \hline
$\chi^2/d.o.f$                           &  264/136                                   & 213/135                                   & 199/134                                     &  197/134                                   & 225/177\\
\hline 
\end{tabular}
\end{center}
\begin{flushleft} 
  \footnotesize{ \hspace{2cm}$^\ast$ Normalization of the $\it{apec}$
    model divided by a solid angle $\Omega$, assumed
    in a uniform-sky ARF\\
    $\hspace{2.2cm}$ calculation (20' radius),
    i.e.\ $\it{Norm} = (1/\Omega) \int n_e n_H dV / (4\pi(1+z)^2)D^2_A)$ cm$^{-5}$ sr$^{-1}$ in unit of \\
    $\hspace{2.2cm} 10^{-14}$, where $D_A$ is the angular diameter distance.\\
    \hspace{2.cm}$^\dagger$ Surface brightness of the {\it power-law} model in the unit of photons s$^{-1}$ cm$^{-2}$ sr$^{-1}$ keV$^{-1}$ at 1 keV\@.} \\
\end{flushleft}
\end{table*}
\renewcommand{\baselinestretch}{1}\selectfont
\renewcommand{\baselinestretch}{1.5}\selectfont
\begin{table*}[htbp]
\begin{center}
\caption{Line centers and surface brightnesses of O\emissiontype{VII} K$\alpha$ and O\emissiontype{VIII} K$\alpha$ emission.}
\label{table:ovii-oviii-norm-center}
\begin{tabular}{lccccc} \hline\hline
                                                                                                                           & ON-FILAMENT                               &  OFFSET-1deg                               & OFFSET-4deg      \\   \hline
\multirow{2}{1.5cm}{O \emissiontype{VII} K$\alpha$}~~center [keV]   & 0.569$\pm$0.006                          &  0.572$^{+0.001}_{-0.002}$        & 0.571$\pm$0.003   \\  
                                    \hspace{2.0cm}SB $^{\ast}$                                     & 11.8$^{+1.6}_{-1.7}$                    &  12.6$^{+0.6}_{-0.7}$                    & 12.9$^{+1.6}_{-1.3}$   \\ \hline
\multirow{2}{1.5cm}{O \emissiontype{VIII} K$\alpha$}~~center [keV]  & 0.654$^{+0.008}_{-0.006}$         & 0.662$^{+0.003}_{-0.004}$         & 0.651$^{+0.005}_{-0.003}$  \\  
                                    \hspace{2.0cm}SB $^{\ast}$                                     & 4.3$\pm$0.7                                   &  2.8$^{+0.2}_{-0.3}$                      & 4.6$\pm$0.6  \\ \hline   \hline 
\end{tabular}
\end{center}
\begin{flushleft} 
\footnotesize{\hspace{3.4cm}$^{\ast}$ Surface brightness in the unit of 10$^{-7}$ photons s$^{-1}$ cm$^{-2}$ arcmin$^{-2}$.}
\end{flushleft}
\end{table*}
\renewcommand{\baselinestretch}{1}\selectfont

\subsection{Emission Analysis in the ON-FILAMENT region}
\label{SEC:emiss-analys-filam}
\begin{figure*}
\begin{minipage}{0.5\hsize}
\begin{center}
\FigureFile(73mm,73mm)
  {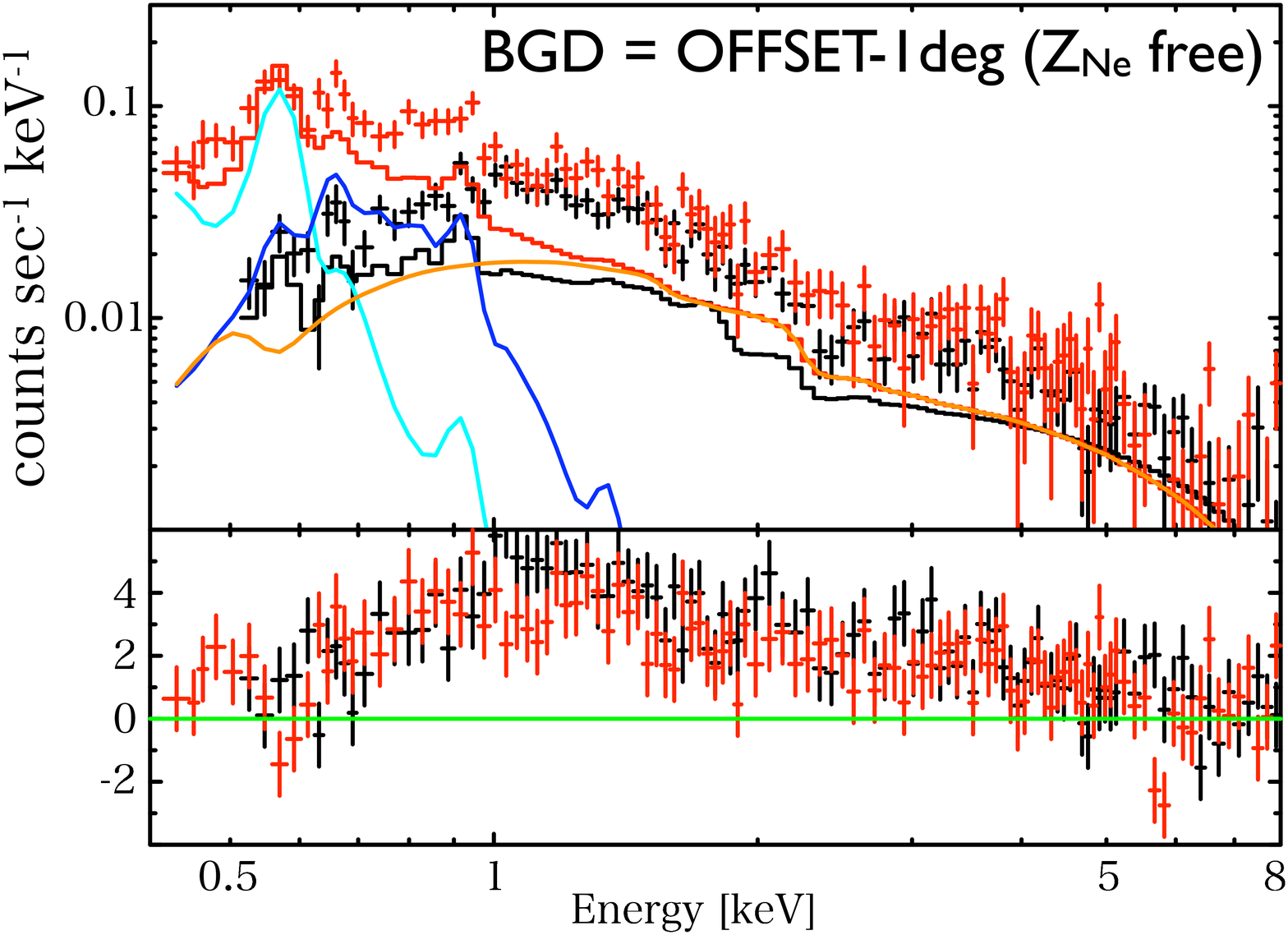}
   \end{center}
   \end{minipage}  
   \begin{minipage}{0.5\hsize}
\begin{center}
\FigureFile(73mm,73mm)
  {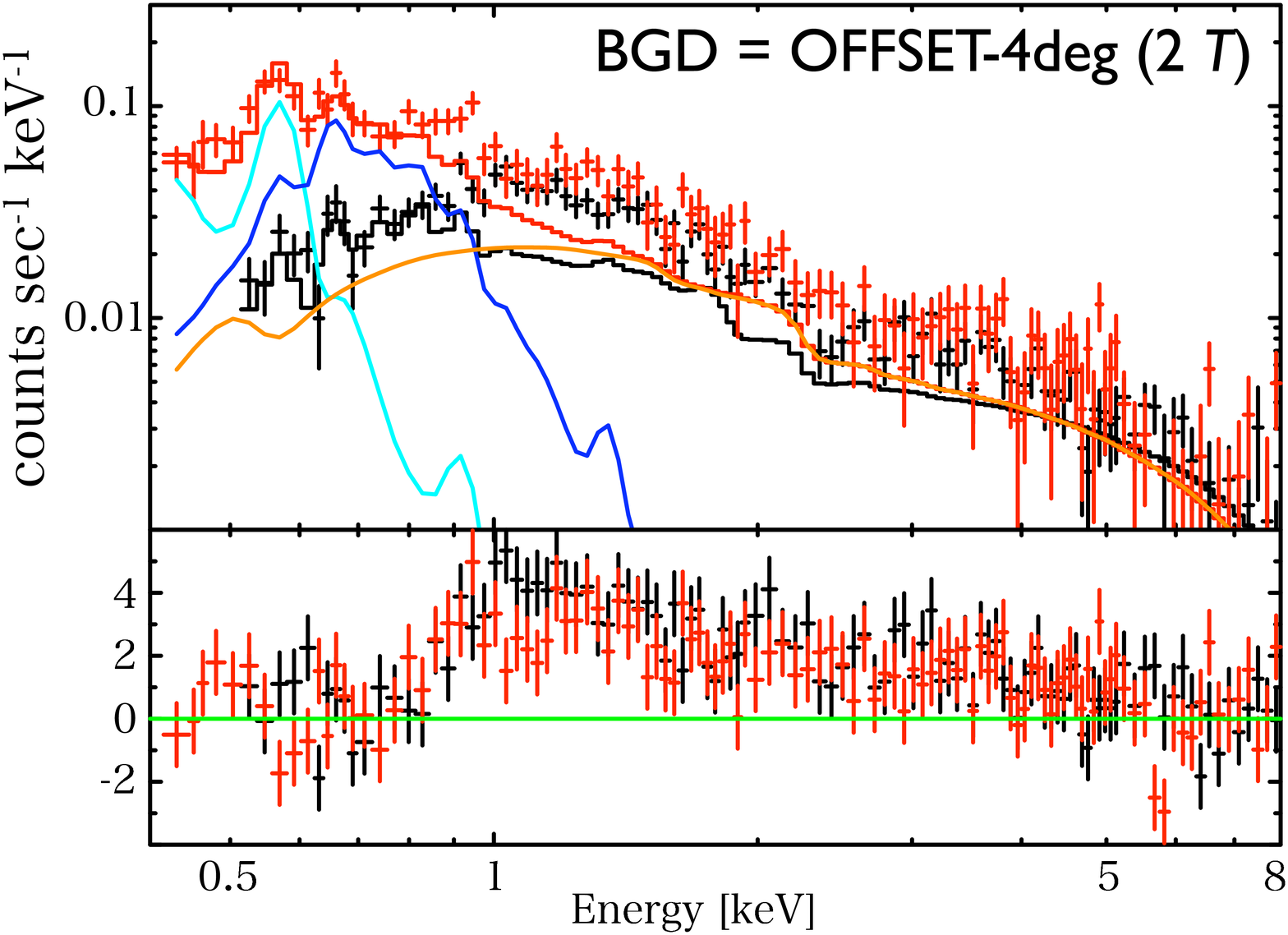}
 \end{center}
 \end{minipage}
 \caption{
Spectra in 0.4--8.0~keV in the ON-FILAMENT region. 
The foreground emission and the CXB are fixed to the best fit parameters 
obtained in the OFFSET-1deg and the OFFSET-4deg regions, respectively. 
Left: fitted with the $Z_\mathrm{Ne}$ free model of the OFFSET-1deg region in Table \ref{table:OFFSET-1deg-2}. 
Right: fitted with the 2 $T$ model of the OFFSET-4deg region in Table \ref{table:OFFSET-1deg-2}.}
   \label{fig:onfilament_offset-1deg-offset-4deg}
   \end{figure*}

In this section, we examined 
whether the ON-FILAMENT can be explained 
by a normal emission model (LHB+SWCX, MWH and CXB) as is the case of the OFFSET-4deg region or 
any excess component is needed as is the case of the OFFSET-1deg region.
As the starting point, we simply compared the obtained spectrum between offset regions and the ON-FILAMENT region 
by fixing all parameters to the best fit values obtained in the OFFSET-1deg ($Z_\mathrm{Ne}$ free in Table \ref{table:OFFSET-1deg-2}) 
and the OFFSET-4deg regions, respectively as shown in Figure \ref{fig:onfilament_offset-1deg-offset-4deg}.
We found that large residuals remained not only in a soft band below 1 keV 
where the Galactic emission is responsible for 
but also in a hard band above 1 keV.

We evaluated the excess emission in the ON-FILAMENT region in comparison with the offset region 
in consideration with field-to-field variation of the background intensity.
The spectra of the ON-FILAMENT region are fitted simultaneously with the background parameters linked together.
We let the normalization of the two Galactic thermal components vary with the same
scaling factor ({\it f}), and normalization of the CXB component was set as a free parameter independent of {\it f}.

\begin{figure*}
\begin{minipage}{0.5\hsize}
\begin{center}
\FigureFile(73mm,73mm)
  {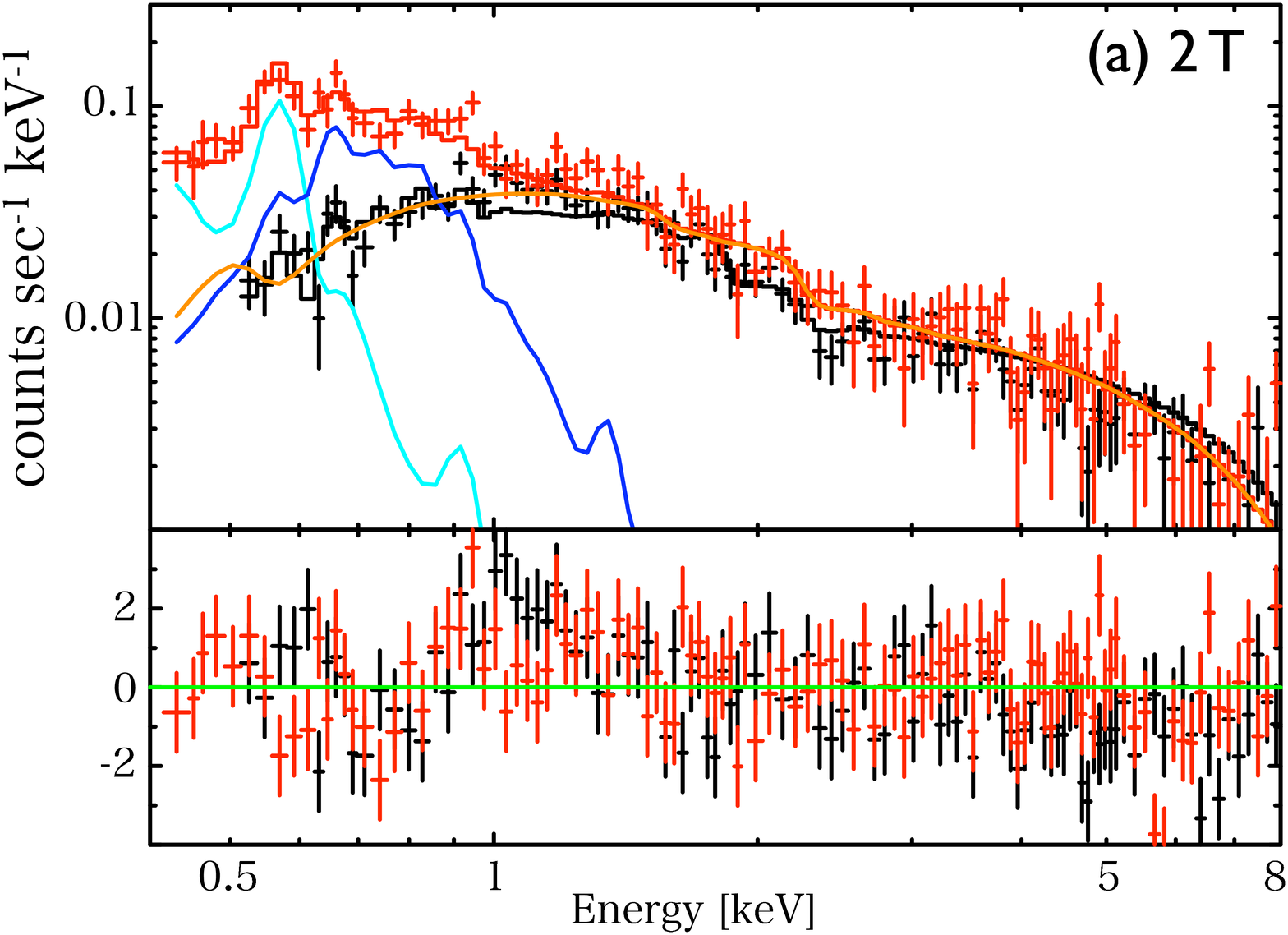}
   \end{center}
   \end{minipage}  
   \begin{minipage}{0.5\hsize}
\begin{center}
\FigureFile(73mm,73mm)
  {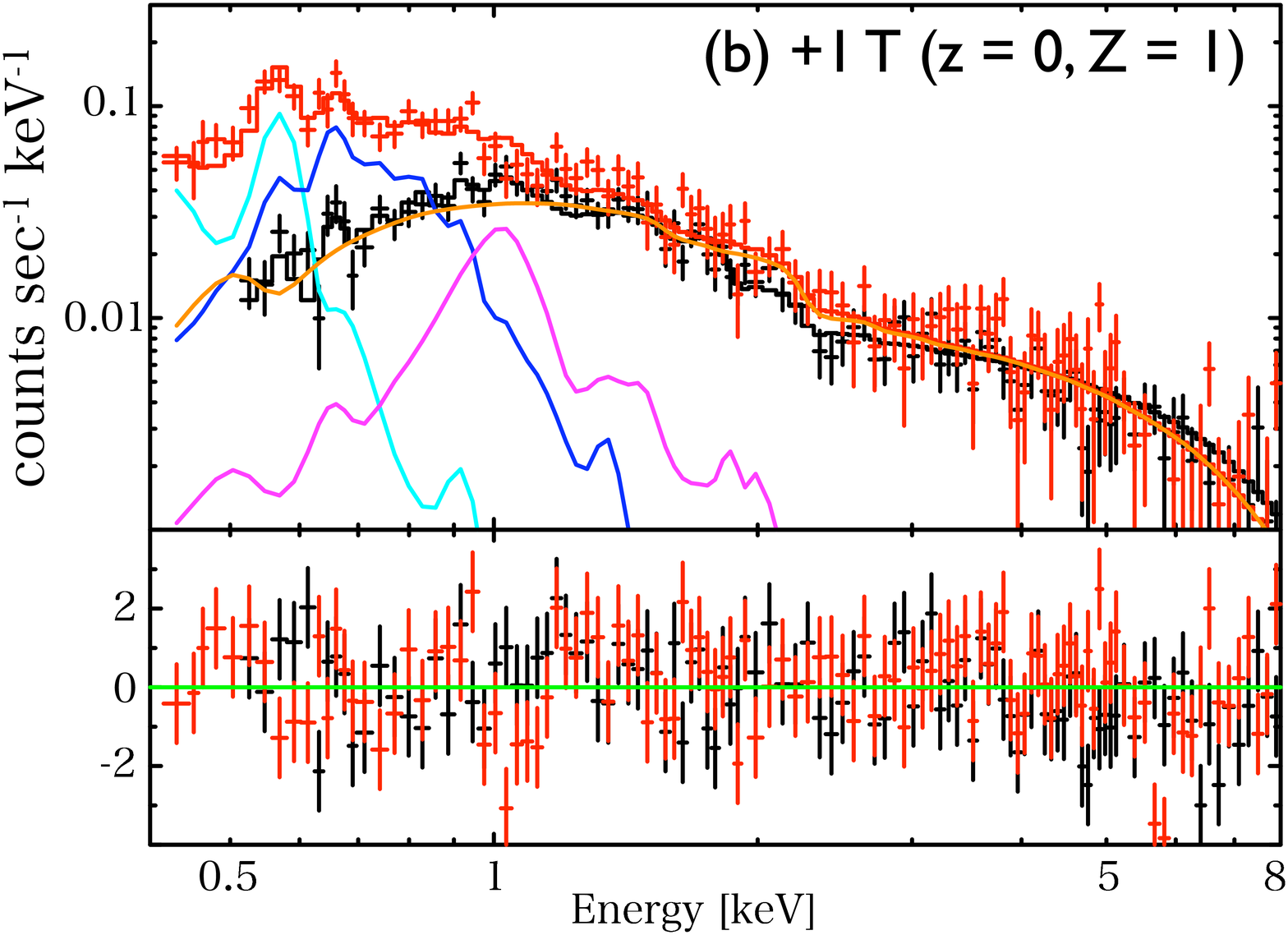}
 \end{center}
 \end{minipage}
   \begin{minipage}{0.5\hsize}
\begin{center}
\FigureFile(73mm,73mm)
  {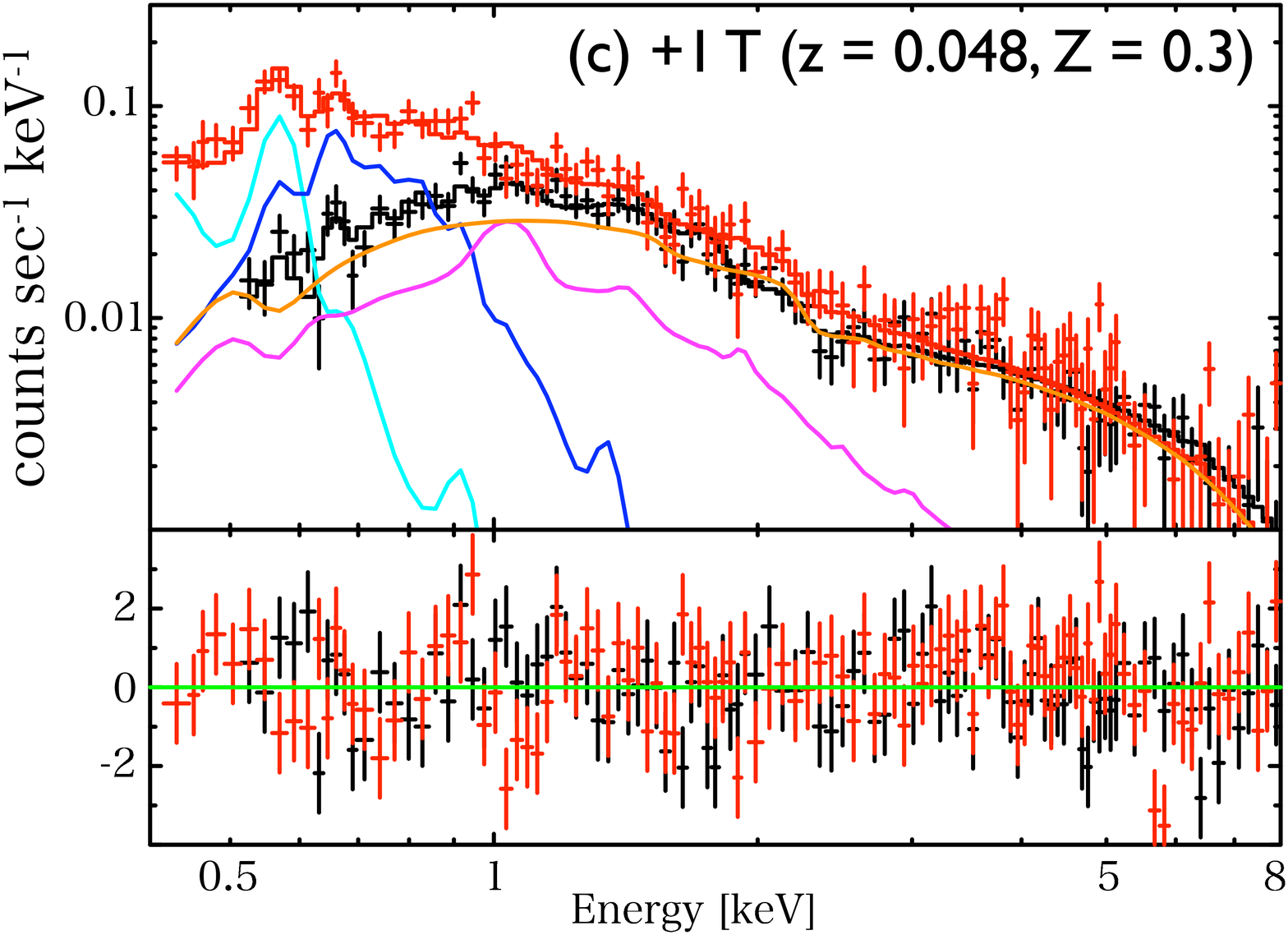}
   \end{center}
   \end{minipage}  
  \begin{minipage}{0.5\hsize}
\begin{center}
\FigureFile(73mm,73mm)
  {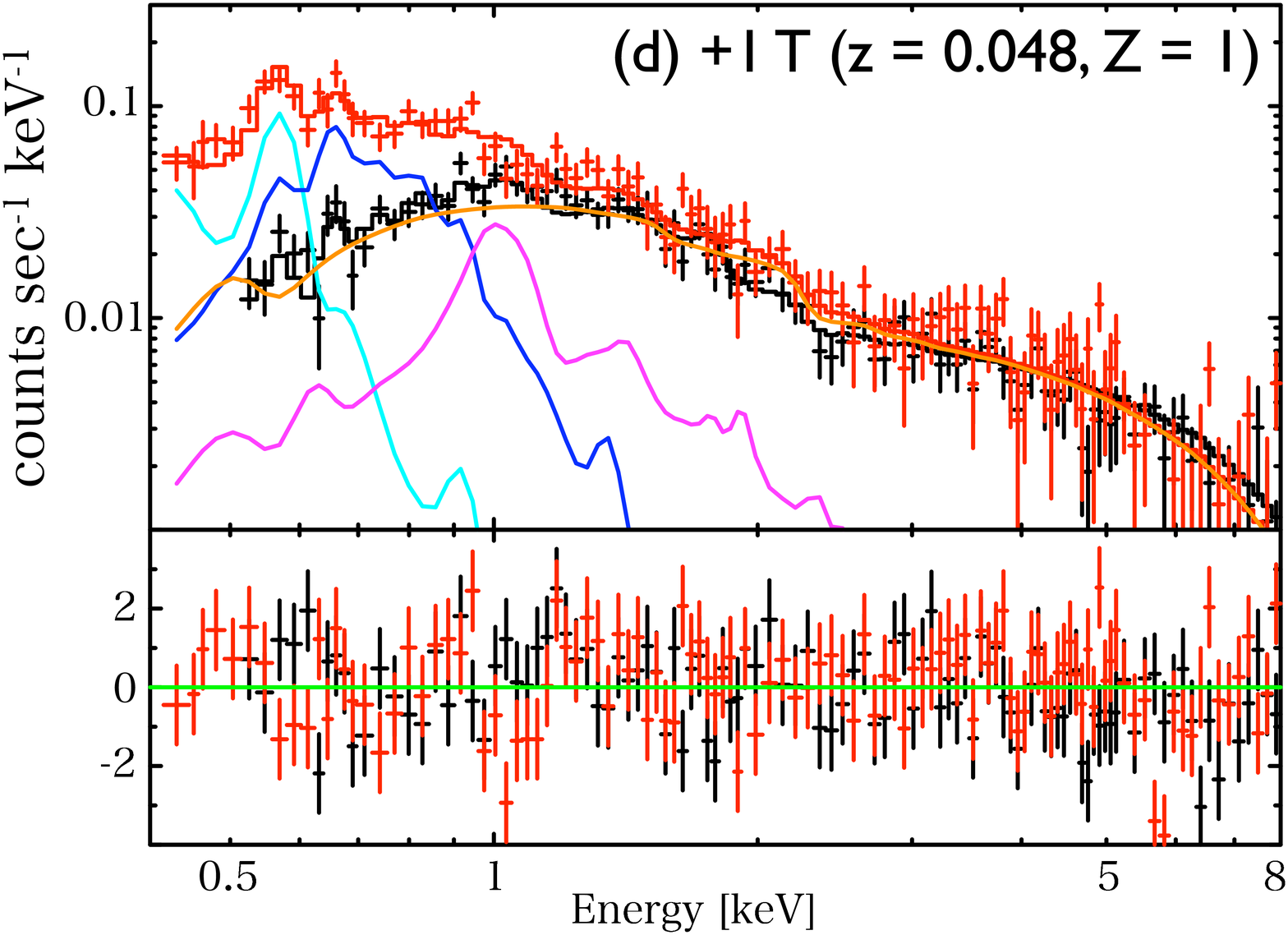}
   \end{center}
   \end{minipage}  
 \caption{
 Spectra in 0.4--8.0~keV in the ON-FILAMENT region. 
The foreground emission is determined 
by fitting with the OFFSET-4deg region (2 $T$ in Table \ref{table:OFFSET-1deg-2}) simultaneously and 
spectra of the OFFSET-4deg are removed in this figure.
Red and black data are for BI and FI sensors, respectively.
The spectral data are the same for the four panels, but the
fitted models and residuals are different.
Top left: fitted with typical foreground/background model,
i.e., the sum of (1), (2) and (3) of Fig.~\ref{fig:OFFSET-1deg-2-2t}.
Top right: fitted with a model with a sum of
(1), (2), (3), and another thin thermal plasma.
The redshift and abundance are $z=0$ and $Z=1$~solar, respectively.
Bottom left: same as top right, but $z=0.048$ and $Z=0.3$~solar.
Bottom right: same as top right, but $z=0.048$ and $Z=1$~solar.
Cyan, blue, orange and magenta lines indicate LHB+SWCX,
   MWH, CXB, and the
   additional higher temperature plasma, respectively.
   The photon index of the CXB is fixed to be 1.4.}
   \label{fig:onfilament_offset-4deg}
   \end{figure*}

Firstly, the OFFSET-4deg region was used as a background spectrum 
because the OFFSET-4deg region was fitted well with the typical model 
for the blank X-ray sky as described in the previous section.
As indicated in Figure \ref{fig:onfilament_offset-4deg} and Table \ref{table:onfilament_offset-4deg}, 
the energy spectrum of the ON-FILAMENT region below 1 keV is well represented by the
typical Galactic emission by the same Galactic emission as OFFSET-4deg with a scaling factor $f$ = 0.98$\pm$0.07.
The CXB surface brightness was 13.3 photons s$^{-1}$ cm$^{-2}$ keV$^{-1}$ at 1 keV, 
which is significantly larger than the expected value from the detection limit of removed point sources 
(8.3 photons s$^{-1}$ cm$^{-2}$ keV$^{-1}$ at 1 keV) if we estimate according to \citet{virial2}.
Note that the fluctuation can be estimated by the same way described in \citet{virial2} 
from Ginga observations \citep{1989PASJ...41..373H} 
to be $\sim$9 \% considering the Poisson noise of the number of sources in the FOV   
under the detection limit. 
This suggests that some sort of continuum emission may exist in addition to the
CXB in the energy range above 1~keV\@.
We thus added one more thermal plasma to evaluate the excess continuum at a redshift of either
0 or that of the supercluster, 
with the abundance of the additional plasma fixed to the solar value. 
The $3T$ model improved the fit compared to the $2T$ model.  
The difference in the goodness of the fit between the two redshifts are
small.  Hence we cannot statistically determine the origin of the
emission.  
These $3T$ fit results are shown and summarized
in Figure \ref{fig:onfilament_offset-4deg} and Table \ref{table:onfilament_offset-4deg}.
However, the CXB level is still high and  
a normalization of the CXB and a temperature of the additional plasma are correlated strongly.
If we restrict the normalization of the CXB up to 
10.5 photons s$^{-1}$ cm$^{-2}$ keV$^{-1}$ sr$^{-1}$ at 1 keV\@ corresponding to 
3 $\sigma$ upper limit of the expected value in the ON-FILAMENT region, 
the temperature of the additional plasma in ($z=0$, $Z=1~Z_\odot$) model of Table \ref{table:onfilament_offset-4deg} 
goes up to 2.1$^{+0.8}_{-0.2}$ keV and 10.2$^{+0.4}_{-0.2}$ in the normalization.
Thus a temperature of the additional plasma can be higher. 
The fit was slightly improved if we fix the abundance of the additional thermal component at 0.3 solar. 
The best-fit temperature and normalization were higher in this case.

\begin{figure*}
\begin{minipage}{0.5\hsize}
\begin{center}
\FigureFile(73mm,73mm)
  {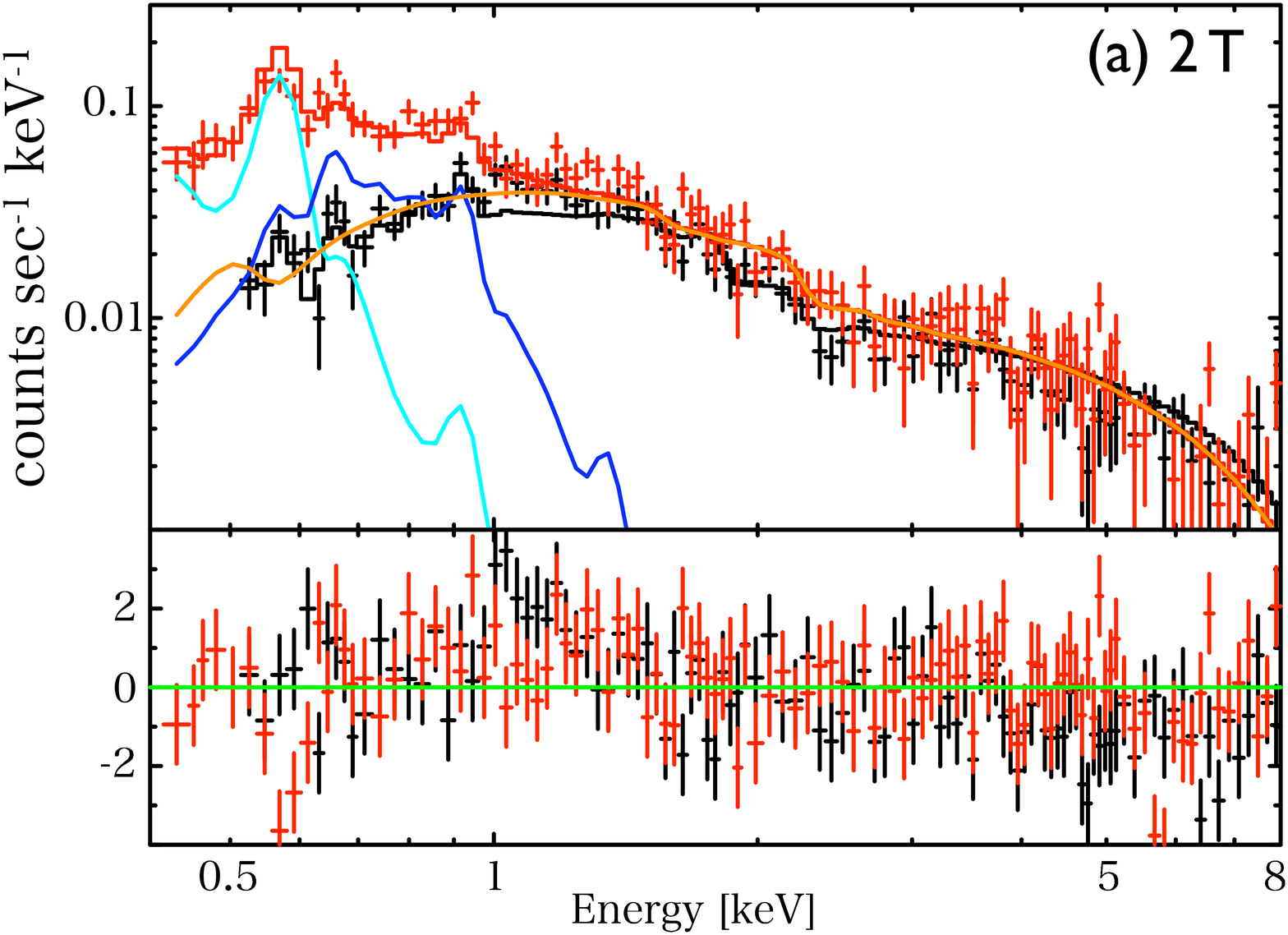}
   \end{center}
   \end{minipage}  
   \begin{minipage}{0.5\hsize}
\begin{center}
\FigureFile(73mm,73mm)
  {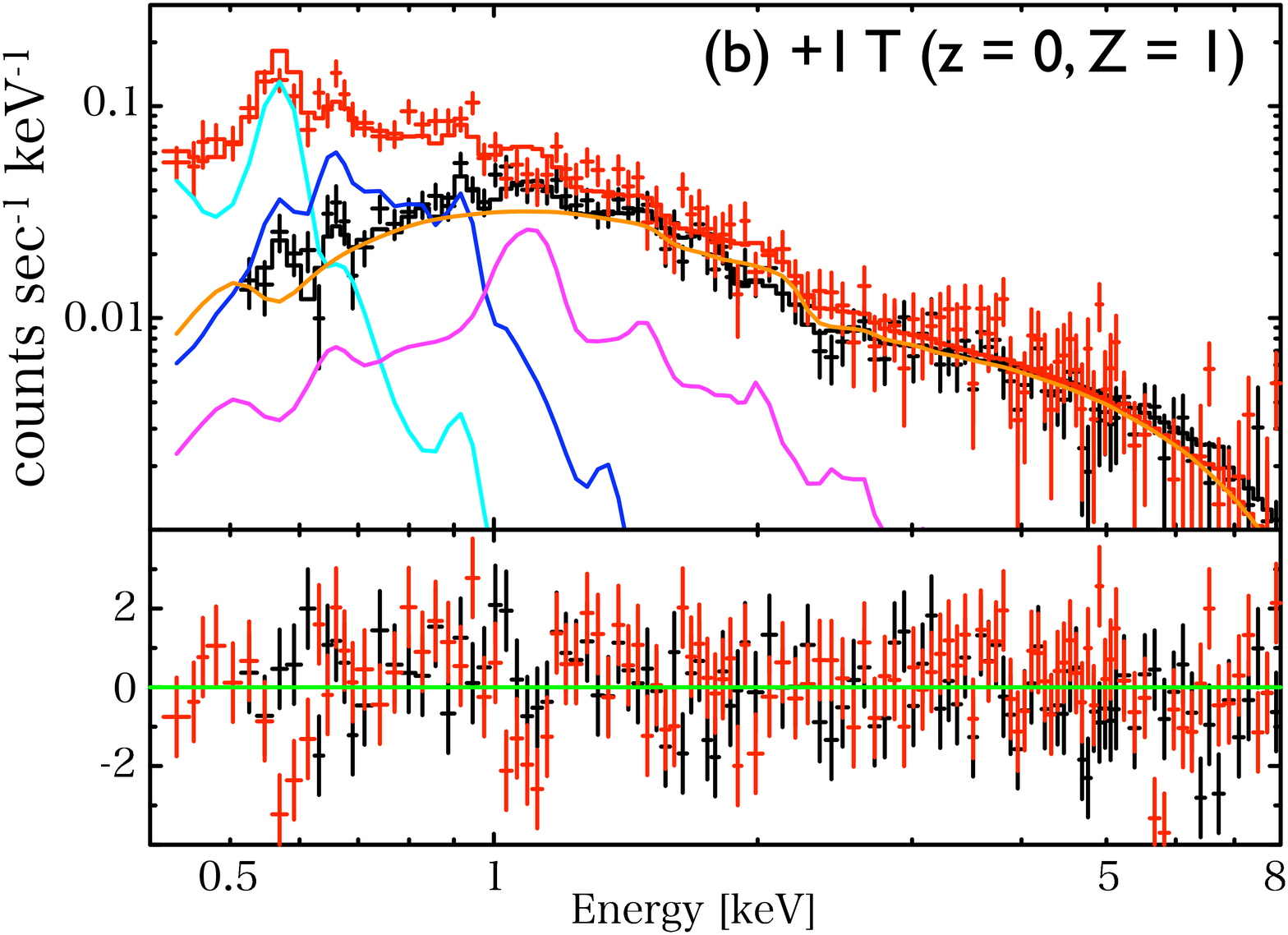}
 \end{center}
 \end{minipage}
   \begin{minipage}{0.5\hsize}
\begin{center}
\FigureFile(73mm,73mm)
  {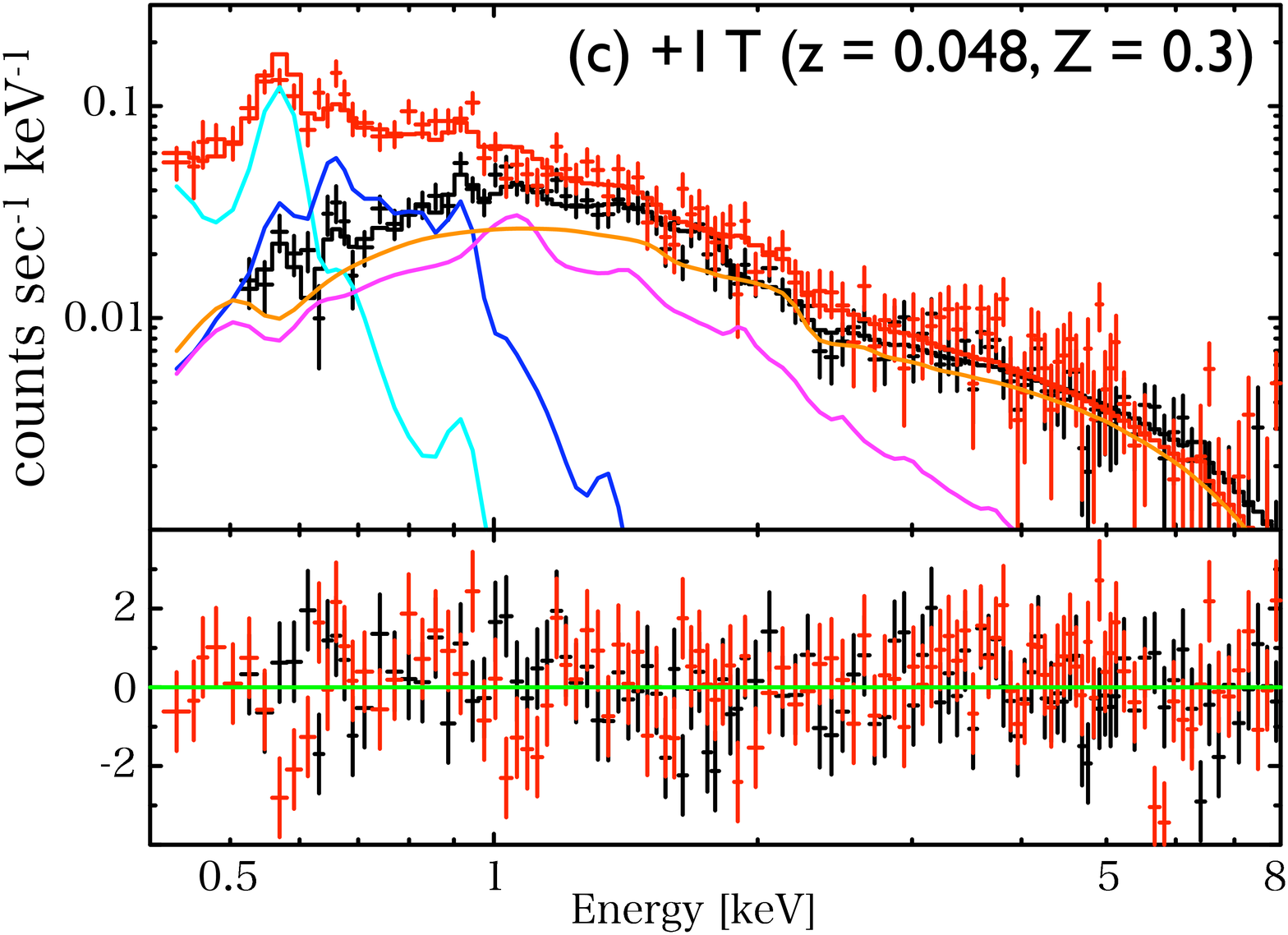}
   \end{center}
   \end{minipage}  
  \begin{minipage}{0.5\hsize}
\begin{center}
\FigureFile(73mm,73mm)
  {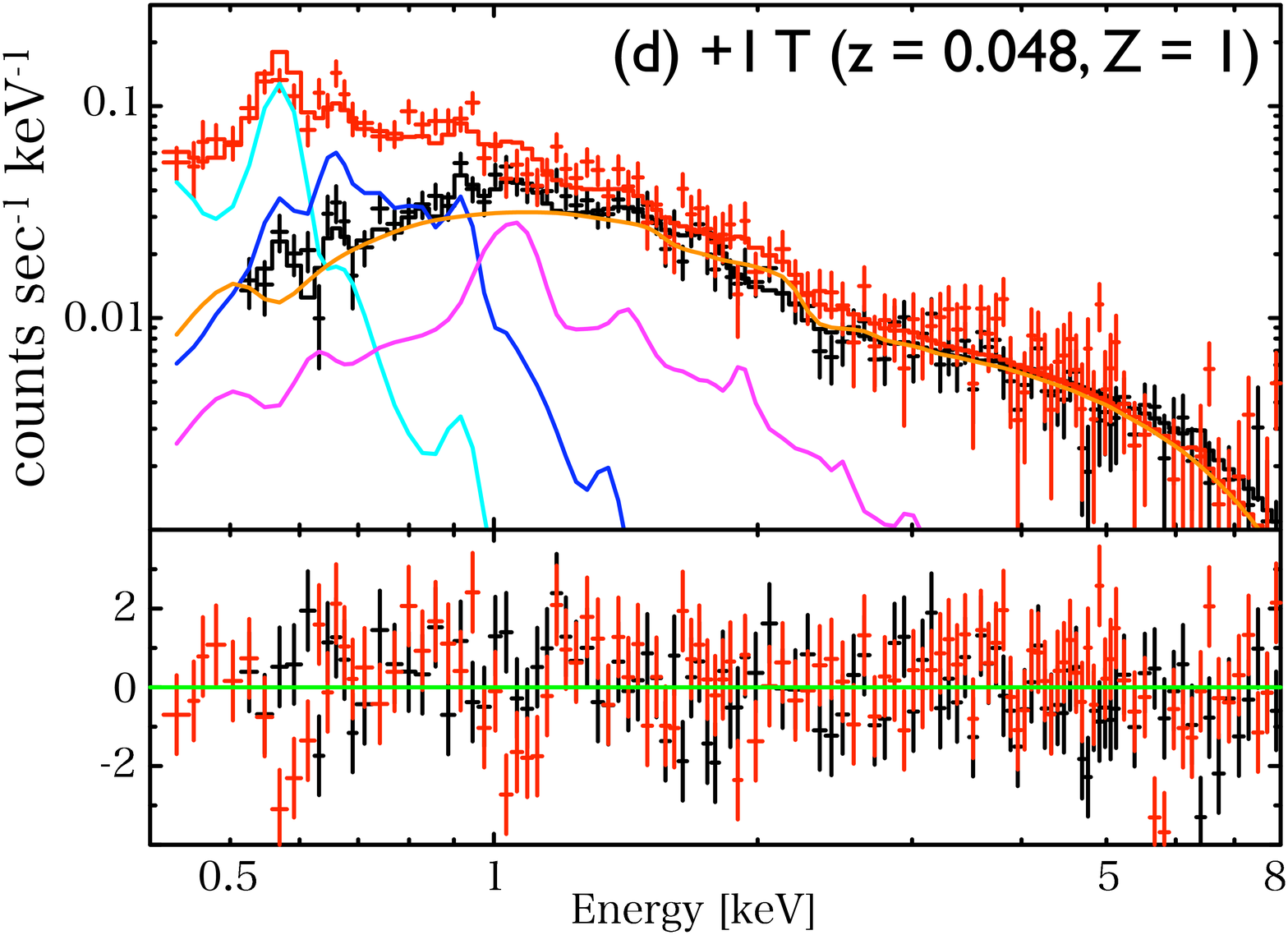}
   \end{center}
   \end{minipage}  
 \caption{The same as Figure \ref{fig:onfilament_offset-4deg}, 
but the $Z_\mathrm{Ne}$ free model of OFFSET-1deg region in Table \ref{table:OFFSET-1deg-2} is uesd as the foreground emission. 
}
   \label{fig:onfilament_offset-1deg}
   \end{figure*}

Next, to evaluate a systematic error caused by the difference of a selected background region, 
the OFFSET-1deg region was also used as a background for the ON-FILAMENT region.
The same analysis was performed as the OFFSET-4deg region and 
spectra and the best fit values are shown and summarized 
in Figure \ref{fig:onfilament_offset-1deg} and Table \ref{table:onfilament_offset1}.
As Galactic emission models, we tried both the two acceptable models 
denoted $Z_\mathrm{Ne}$ free and 3 $T$ ($z=0$) in Table \ref{table:OFFSET-1deg-2}. 
Here, we showed only results of the $Z_\mathrm{Ne}$ free model because 
we confirmed that resulting parameters for the excess emission in the ON-FILAMENT spectrum 
were consistent with each other within the statistical errors.
Resultant parameters for the additional plasma are also consistent with each other 
between the OFFSET-1deg and the OFFSET-4deg.
It means that this kind of systematic error does not affect our results discussed in later section.

We also evaluated line centers and the surface brightness of O
\emissiontype{VII} K$\alpha$ and O \emissiontype{VIII} K$\alpha$
emission lines in the similar manner as we did for offset regions.
The best-fit values are shown in Table
\ref{table:ovii-oviii-norm-center}.
The center energies in the ON-FILAMENT region are again consistent with
zero redshift within the statistical error and the
typical energy determination accuracy of the {\it Suzaku} XIS\@.
\renewcommand{\baselinestretch}{1.5}\selectfont
\begin{table*}[htbp]
\begin{center}
\caption{Best fit parameters of the ON-FILAMENT region modeling the foreground by 
fitting with the OFFSET-4deg region simultaneously.}
\label{table:onfilament_offset-4deg}
\begin{tabular}{ccccccc} \hline\hline
                                                   & 2 $T$
     &\multicolumn{3}{c}{3 $T$}
 \\ 
&& $z=0$, $Z=1~Z_\odot$ & $z=0.048$, $Z=1~Z_\odot$ & $z=0.048$, $Z=0.3~Z_\odot$\\
\hline                      
{\it f}                                          & 0.98$\pm$0.07                        & 0.90$\pm$0.07                      & 0.91$\pm$0.07                      & 0.87$\pm$0.07\\
$kT_1$ [keV]                           & 0.093$^{+0.012}_{-0.006}$ & 0.091$\pm$0.010                  & 0.091$\pm$0.010                 & 0.091$^{+0.011}_{-0.008}$   \\
$\it{Norm_1}$$^{\ast}$/{\it f} & 100.9$^{+38.3}_{-38.2}$      & 110.9$^{+68.5}_{-39.2}$      & 110.0$^{+67.6}_{-39.2}$     & 109.0$^{+56.9}_{-39.7}$   \\
$kT_2$ [keV]                           & 0.250$^{+0.019}_{-0.014}$ & 0.235$^{+0.015}_{-0.012}$ & 0.234$^{+0.015}_{-0.012}$ & 0.235$^{+0.016}_{-0.012}$   \\
$Norm_2$$^{\ast}$/{\it f}       & 14.3$\pm$2.4                         & 16.0$\pm$2.5                         & 16.0$\pm$2.5                         & 15.9$\pm$2.5      \\
$\Gamma$ (fix)                       & 1.4 (fix)                                     & 1.4 (fix)                                    & 1.4 (fix)                                     &  1.4 (fix)                           \\
 $\it{SB}$$^\dagger$             &13.3$\pm$0.3                          & 12.0$\pm$0.5                         & 11.7$\pm$0.6                         & 10.0$^{+1.4}_{-1.7}$        \\ \hline
$kT_3$                                     &                                                   & 1.3$^{+0.3}_{-0.2}$              & 1.6$^{+0.6}_{-0.3}$               & 2.0$^{+0.8}_{-0.5}$   \\
Redshift                                   &                                                   & 0                                                & 0.048 (fix)                                & 0.048 (fix)                      \\
$Z~(Z_{\odot})$                      &                                                   & 1.0 (fix)                                    & 1.0 (fix)                                     & 0.3 (fix)                         \\
$Norm_3$$^{\ast}$               &                                                   & 3.6$\pm$1.0                           & 6.6$^{+5.3}_{-3.1}$               & 22.7$^{+8.4}_{-8.3}$  \\ \hline
$\chi^2/d.o.f$                          &  568/398                                  & 497/396                                   & 496/396                                   & 481/396                        \\ \hline 
\end{tabular}
\begin{minipage}[t]{0.8\textwidth}
\begin{flushleft} 
  \footnotesize{ $^\ast$ Normalization of the
      $\it{apec}$ models.  See notes in Table 2 for the definition of
      the normalization. The foreground emission is determined 
      by fitting with the OFFSET-4deg region simultaneously and 
      their normalizations are rescaled by the factor of {\it f}.  \\ 
      $^\dagger$ Surface brightness of the {\it power-law}
    model in the unit of photons s$^{-1}$ cm$^{-2}$ keV$^{-1}$
    sr$^{-1}$ at 1 keV\@.} \\
\end{flushleft}
\end{minipage}
\end{center}
\end{table*}
\renewcommand{\baselinestretch}{1}\selectfont
\renewcommand{\baselinestretch}{1.5}\selectfont
\begin{table*}[htbp]
\begin{center}
\caption{Best fit parameters of the ON-FILAMENT region modeling the foreground by 
fitting with the OFFSET-1deg region simultaneously.}
\label{table:onfilament_offset1}
\begin{tabular}{ccccccc} \hline\hline
                                                   & 2 $T$
     &\multicolumn{3}{c}{3 $T$}
 \\ 
                                                 &                                                   & $z=0$, $Z=1~Z_\odot$           & $z=0.048$, $Z=1~Z_\odot$ & $z=0.048$, $Z=0.3~Z_\odot$\\
\hline                      
{\it f}                                         & 1.24$\pm$0.08                       & 1.18$\pm$0.08                      & 1.16$\pm$0.08                        & 1.10$\pm$0.08\\
$kT_1$ [keV]                         & 0.101$^{+0.011}_{-0.007}$ & 0.100$^{+0.010}_{-0.006}$& 0.100$^{+0.010}_{-0.006}$   & 0.100$^{+0.010}_{-0.006}$   \\
$\it{Norm_1}$$^{\ast}$/{\it f}& 72.2$^{+21.5}_{-20.8}$       & 74.5$^{+21.7}_{-20.4}$        & 75.3$^{+21.8}_{-20.8}$         & 74.6$^{+21.9}_{-20.3}$   \\
$kT_2$ [keV]                         & 0.238$^{+0.014}_{-0.011}$ & 0.231$^{+0.013}_{-0.010}$ & 0.229$^{+0.012}_{-0.010}$  & 0.229$^{+0.012}_{-0.010}$   \\
$Z_\mathrm{Ne}$ (Z$_{\odot}$)& 2.2$^{+0.3}_{-0.2}$ & 2.1$^{+0.3}_{-0.2}$        & 2.1$\pm$0.3                            & 2.1$\pm$0.3\\
$Norm_2$$^{\ast}$/{\it f}     &8.8$^{+1.3}_{-1.7}$                & 9.4$\pm$1.3                           & 9.6$\pm$1.3                             & 9.5$\pm$1.3      \\
$\Gamma$ (fix)                     & 1.4 (fix)                                    & 1.4 (fix)                                      &  1.4 (fix)                                    & \\
 $\it{SB}$$^\dagger$           & 13.5$\pm$0.4                         & 11.0$^{+1.3}_{-1.5}$            & 10.9$^{+0.9}_{-1.3}$               & 9.1$^{+1.4}_{-1.9}$        \\ \hline
$kT_3$                                   &                                                  & 2.1$^{+0.9}_{-1.2}$              & 2.1$^{+0.8}_{-0.5}$                  & 2.3$^{+0.9}_{-0.6}$   \\
Redshift                                  &                                                  & 0                                                & 0.048 (fix)                                 & 0.048 (fix)                      \\
$Z~(Z_{\odot})$                    &                                                   & 1.0 (fix)                                    & 1.0 (fix)                                      & 0.3 (fix)                         \\
$Norm_3$$^{\ast}$             &                                                   & 9.2$^{+6.2}_{-6.8}$               & 11.1$^{+5.9}_{-4.8}$              & 27.5$^{+8.5}_{-7.7}$  \\ \hline
$\chi^2/d.o.f$                        & 570/356                                   & 499/354                                   & 497/354                                    & 475/354                        \\ \hline 
\end{tabular}
\begin{minipage}[t]{0.8\textwidth}
\begin{flushleft} 
  \footnotesize{ $^\ast$ Normalization of the
      $\it{apec}$ models.  See notes in Table 2 for the definition of
      the normalization.  The foreground emission is determined 
      by fitting with the OFFSET-1deg region ($Z_\mathrm{Ne}$ free model in Table \ref{table:OFFSET-1deg-2}) 
      simultaneously and their normalizations are rescaled by the factor of {\it f}. \\ 
      $^\dagger$ Surface brightness of the {\it power-law}
    model in the unit of photons s$^{-1}$ cm$^{-2}$ keV$^{-1}$
    sr$^{-1}$ at 1 keV\@.} \\
\end{flushleft}
\end{minipage}
\end{center}
\end{table*}
\renewcommand{\baselinestretch}{1}\selectfont
\subsection{Upper limit of intensities and densities of
the redshifted component}\label{SEC:upper-limit-intens}


We evaluated the $2\sigma$ upper limit of WHIM density at
$kT= 0.2-0.3$~keV, using upper limits of redshfited
O\emissiontype{VII} K$\alpha$ and O \emissiontype{VIII} K$\alpha$
emission line intensities, with the same method as
\citet{intensity-overdensity}.
To determine the upper limits, we added in the model described in
\S~\ref{SEC:analysis--results} two Gaussian functions that represent
redshifted O \emissiontype{VII} K$\alpha$ and O \emissiontype{VIII}
K$\alpha$ emission lines, i.e., we used the following models: {\it
  apec$_\mathrm{1}$} + {\it phabs}$\times$ ({\it vapec$_\mathrm{2}$}+
{\it power-law}) for the OFFSET-1deg region, and {\it
  apec$_\mathrm{1}$} + {\it phabs}$\times$ ({\it apec$_\mathrm{2}$}+
{\it apec$_\mathrm{3}$}+ {\it power-law}) for the ON-FILAMENT region.
Temperature of the $\it{apec}$ models was fixed to the best fit values
and the energy of the O lines were set to be the redshifted values
(0.544 keV for O \emissiontype{VII} K$\alpha$ and 0.624 keV for O
\emissiontype{VIII} K$\alpha$).  We assume the velocity dispersion of
the line is negligible and the widths of the Gaussian were set to 0.
The upper limits of 
O \emissiontype{VII} K$\alpha$ and O \emissiontype{VIII}
K$\alpha$ lines for the OFFSET-1deg region are, respectively,
$4.2 \times 10^{-7}$ and $4.0 \times 10^{-8}$
photons s$^{-1}$ cm$^{-2}$ arcmin$^{-2}$.
For the ON-FILAMENT region, 
they are $2.4$ and $1.3 \times 10^{-7}$ photons s$^{-1}$
cm$^{-2}$ arcmin$^{-2}$ with the OFFSET-1deg background template and 
$9.0$ and $1.5 \times 10^{-7}$ photons s$^{-1}$ cm$^{-2}$ arcmin$^{-2}$ 
with the OFFSET-4deg template.

As described in \citet{intensity-overdensity}, we constrain the density
of the WHIM assuming a uniform density and temperature distribution,
as a function of the temperature.  We adopted $3\times 10^{6}$ K which
gave the maximum emissivity of O \emissiontype{VIII} K$\alpha$ line.

The 2$\sigma$ upper limit intensity of O \emissiontype{VIII} K$\alpha$ in the
ON-FILAMENT region with the OFFSET-4deg background template, 
$I< 1.5\times 10^{-7}$ photons s$^{-1}$ cm$^{-2}$
arcmin$^{-2}$, gives a constraint on the gas density.  We assumed the
line-of-sight length $L=3$~Mpc, which is about the same as the
distance between A3556 and A3558.  The density is estimated as
\begin{equation}
n_H < 7.7\times10^{-5} {\rm~cm^{-3}}~
\left(\frac{Z}{0.1~Z_{\odot}}\right)^{-1/2} 
\left(\frac{L}{3~{\rm Mpc}}\right)^{-1/2}.
\end{equation}
Here an electron-to-hydrogen number density ratio $n_e/n_H$ of 1.2 
is assumed because H and He are fully ionized.
The overdensity $\delta\equiv n_H/\bar{n}_H -1$ is
calculated to be 
\begin{equation}
\delta < 380~
\left(\frac{Z}{0.1~Z_{\odot}}\right)^{-1/2} 
\left(\frac{L}{3~{\rm Mpc}}\right)^{-1/2}.
\end{equation}

We can also constrain the density of the redshifted thermal plasma,
i.e., the supercluster plasma, from the
  emission measure of the continuum emission as discussed in \S~4.1.  
  To extract the upper limit of the density of
  the supercluster plasma, we assumed all of the observed
  excess emission to be due to the supercluster plasma. 
From the definition of the $\it{norm}$ in the
$\it{apec}$ model,
\begin{equation}
n_H < 1.1\times10^{-5}
\left({\it norm}\right)^{1/2}
\left(\frac{L}{3~{\rm Mpc}}\right)^{-1/2}.
\end{equation}
With {\it norm} = 1.8, we obtained $1.5\times 10^{-5}\ (L/3~{\rm
  Mpc})^{-1/2}$ cm$^{-3}$ and $\delta<70\ (L/3~{\rm Mpc})^{-1/2}$ for the
OFFSET-1deg region. For the ON-FILAMENT region with the OFFSET-1deg background template, 
{\it norm} = 37.9 indicated $6.8\times 10^{-5}\ (L/3~{\rm Mpc})^{-1/2}$ cm$^{-3}$ and
$\delta< 330 \ (L/3~{\rm Mpc})^{-1/2}$ .  The corresponding continuum
flux $F$ in the ON-FILMENT region in 0.5--2.0~keV is $F = 1.3^{+0.4}_{-0.3}
\times 10^{-15}~\mathrm{ergs~ s^{-1}~ cm^{-2}~ arcmin^{-2}}$ with the OFFSET-1deg background template.
The obtained flux is consistent with that with the OFFSET-4deg background template.
Resultant 2$\sigma$ upper limits of $n_H$ and $\delta$ are summarized in Table
\ref{table:intensity-overdensity-upperlimits}.  In the case of O
\emissiontype{VII} K$\alpha$, we assumed $2\times 10^{6}$ K
corresponding to the maximum emissivity of O \emissiontype{VII}
K$\alpha$ lines.
\renewcommand{\baselinestretch}{1.5}\selectfont
\begin{table*}[htbp]
\begin{center}
\caption{2$\sigma$ upper limit densities and overdensities using the redshifted O \emissiontype{VII} K$\alpha$, 
O \emissiontype{VIII} K$\alpha$ and, continuum emissions.}
\label{table:intensity-overdensity-upperlimits}
\begin{tabular}{llcccc} \hline\hline
REGION &                                                                                                                   & \multicolumn{2}{c}{ON-FILAMENT}  & OFFSET-1deg   \\ \hline
\multicolumn{2}{l}{O \emissiontype{VII}~($T$ = 2$\times$10$^{6}$ K,  $Z=0.1~Z_\odot$ and, $L=3$~Mpc)} & BGD=OFFSET-1deg$^{\ast}$ & BGD=OFFSET-4deg$^{\dagger}$ &        \\   \hline
SB$$             & 10$^{-7}$ photons s$^{-1}$ cm$^{-2}$ arcmin$^{-2}$            & 2.4                              & 9.0                      &   4.2                  \\  
$n_H$            & 10$^{-5}$ cm$^{-3}$                                                                     & 7.9                              & 15                      &   10                        \\
$\delta$             &                                                                                                        & 390                            & 750                      &   510          \\ \hline
\multicolumn{2}{l}{O \emissiontype{VIII}~($T$ = 3$\times$10$^{6}$ K,  $Z=0.1~Z_\odot$ and, $L=3$~Mpc)}  & & &                                  \\ \hline
SB                & 10$^{-7}$ photons s$^{-1}$ cm$^{-2}$ arcmin$^{-2}$              & 1.3                             &  1.5                      &  0.4 \\  
$n_H$             & 10$^{-5}$ cm$^{-3}$                                                                    & 7.3                             & 7.7                       &  4.0     \\ 
$\delta$ &                                                                                                                    & 360                           &  380                      &  200       \\ \hline
\multicolumn{2}{l}{Continuum~($L=3$~Mpc)}                                                     &                                    &                           &                             \\ \hline
$n_H^{\ddagger}$        & 10$^{-5}$ cm$^{-3}$                                                    & 6.8                              & 6.4                    & 1.5       \\ 
$\delta^{\ddagger}$         &                                                                                       & 330                             & 310                 & 70        \\ \hline \hline
\end{tabular}
\end{center}
\begin{flushleft} 
\footnotesize{$^{\ast}$ The ($Z_\mathrm{Ne}$ free) column in Table \ref{table:OFFSET-1deg-2}.\\
$^{\dagger}$ The Nominal column in Table \ref{table:OFFSET-1deg-2} right.\\
$^{\ddagger}$ Extracted by the $\it{Norm}$ value 
in the $\it{apec}$ model of the redshifted thermal plasma of the ($z=0.048$, $Z=0.3~Z_\odot$) column 
in Table \ref{table:onfilament_offset-4deg} and \ref{table:onfilament_offset1} right.}
\end{flushleft}
\end{table*}
\renewcommand{\baselinestretch}{1}\selectfont
\section{Discussion}
\subsection{Origin of the redshifted thermal emission}
\label{SEC:orig-therm}

As reported in \S~\ref{SEC:introduction}, excess emission compared
to the surrounding background level was detected with {\it ROSAT}
(\cite{rosat}) in 0.1--2.4 keV\@. The location was between A3558 and
A3556, namely the same region as in the present Suzaku study.
The intensity of this excess emission in 0.1--2.4~keV is
$2\times 10^{-15}$ ergs s$^{-1}$ cm$^{-2}$ arcmin$^{-2}$.
The intensities of the additional redshifted thermal component in the
{\it Suzaku} spectrum are $1.9^{+0.6}_{-0.5} \times 10^{-15}$ ergs
s$^{-1}$ cm$^{-2}$ arcmin$^{-2}$ and $1.6^{+0.6}_{-0.5} \times 10^{-15}$ ergs
s$^{-1}$ cm$^{-2}$ arcmin$^{-2}$ with the OFFSET-1deg and 
the OFFSET-4deg background templates, respectively. 
This is consistent with the reported excess with {\it ROSAT}\@ within the statistical error. 

\begin{figure*}
\begin{minipage}{0.5\hsize}
\begin{center}
\FigureFile(80mm,80mm)
  {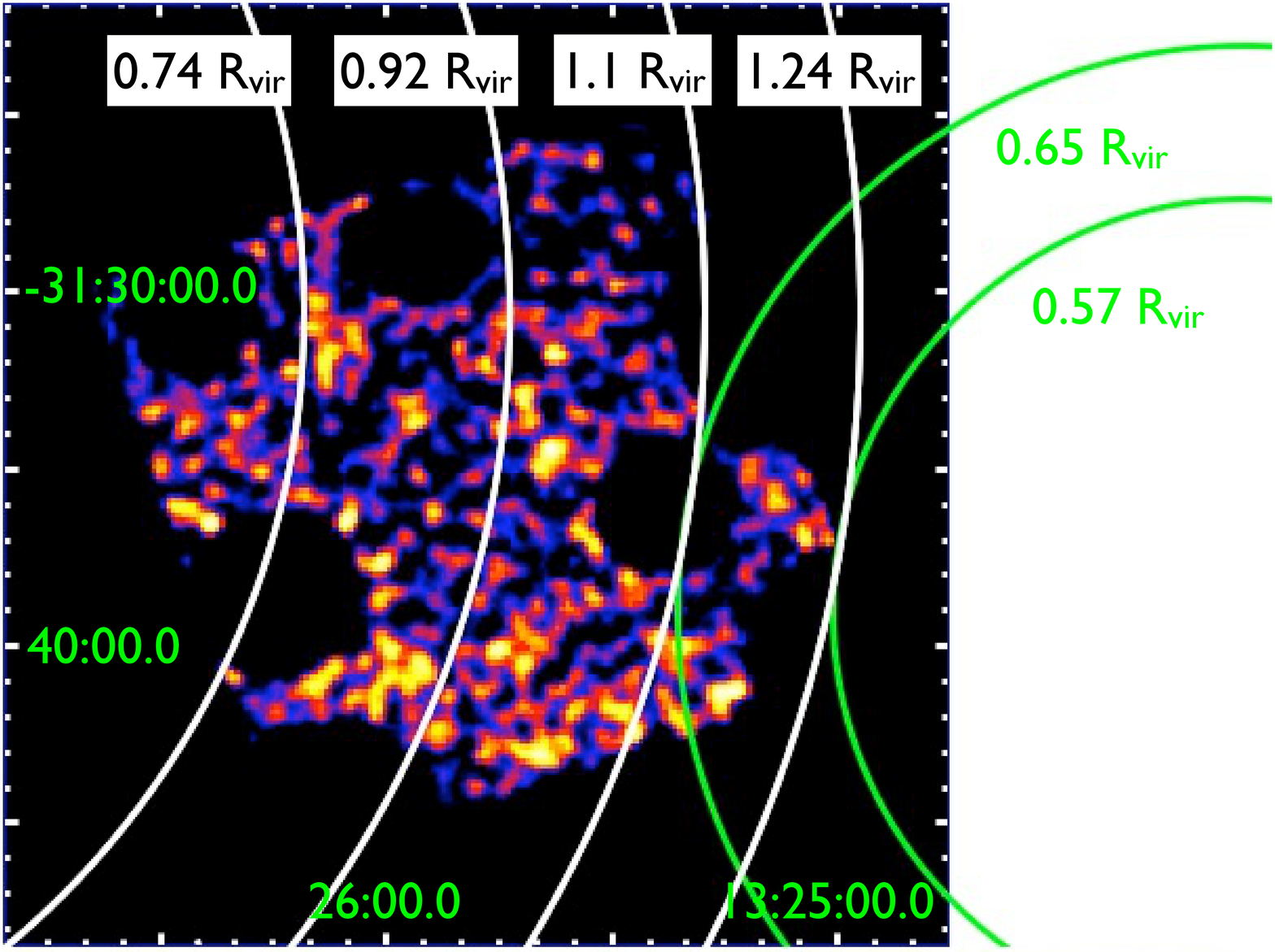}
   \end{center}
   \end{minipage}  
   \begin{minipage}{0.5\hsize}
\begin{center}
\FigureFile(80mm,80mm)
  {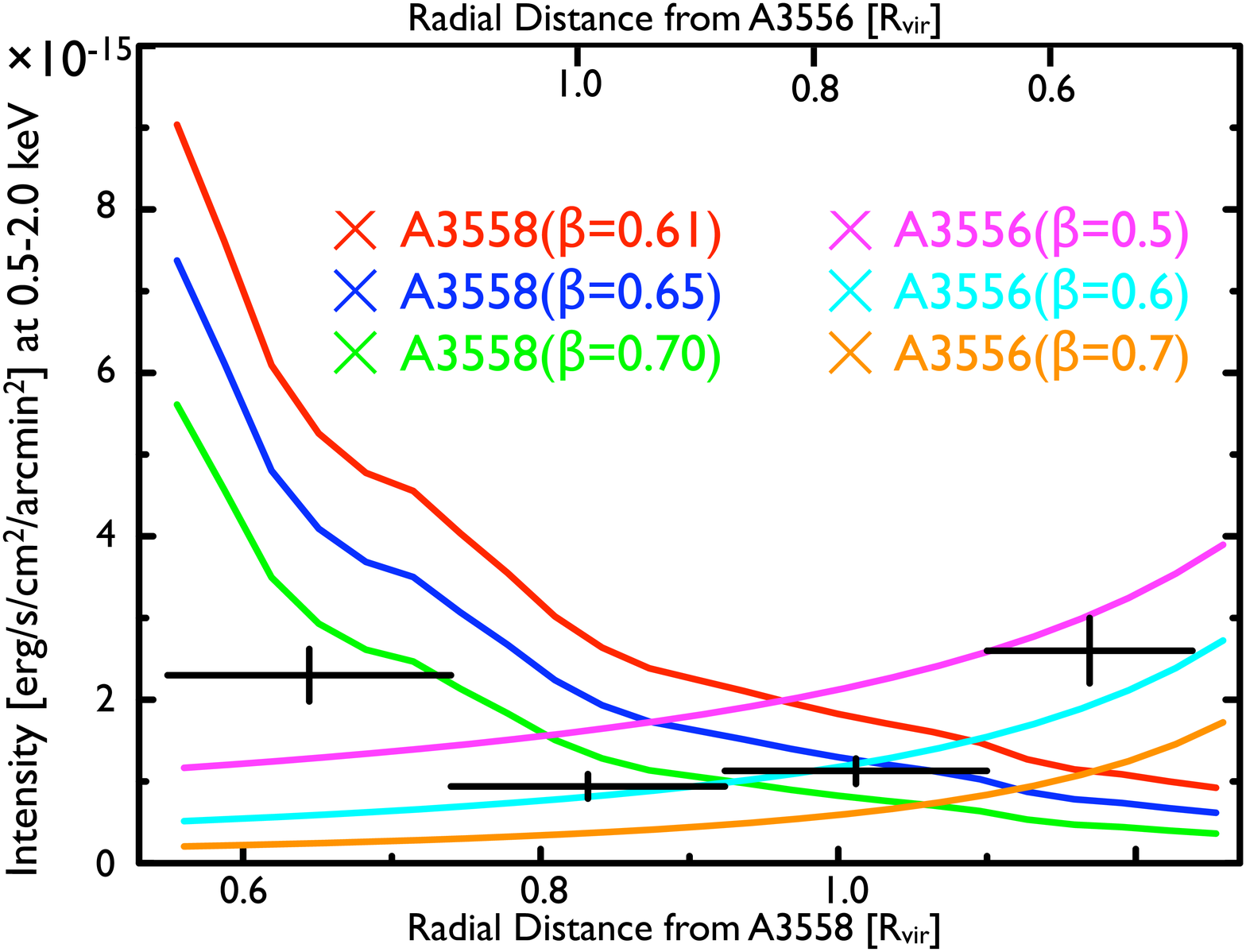}
 \end{center}
 \end{minipage}
 \caption{Left: Regions for the spectral analysis are shown on the XIS
   BI image with radial distances from A3558 (white) and A3556
   (green).  Right: Radial profile of the observed excess emission
   compared with the expected ones.  Black symbols show the observed
   excess.  Curves falling to the right are expected profiles of A3558
   assuming $\beta = 0.61$ (red), 0.65 (blue) and 0.70 (green),
   respectively. Other curves show brightness profiles of A3556
   assuming $\beta = 0.5$ (magenta), 0.6 (cyan) and 0.7 (orange),
   respectively.}
    \label{fig:radial-profile-flux-comparison}
   \end{figure*}

The possible origins of this excess emission are (1) cluster emission
at the outskirt region, (2) stray light from the inner bright cluster
emission, (3) unresolved point sources and (4) plasma associated with
the supercluster.  First, we focused on the first two possibilities.
According to \cite{a3558-sb}, surface brightness distribution
of bright cluster A3558 within $< 0.55\ (R_\mathrm{vir}$) shows a
roughly elliptical shape with a superposition of 3 components: the
A3558 cluster emission (an elliptical King model with $\beta = 0.61$),
the central galaxy emission (approximated by a gaussian) and the
spatially constant background.
This $\beta$ value is about $20\%$ higher than that of ASCA result by
\cite{fukazawa-2004} obtained by fitting with a spherically
symmetric $\beta$ model.  Therefore, if observed excess emission by
{\it Suzaku} is associated with A3558, the intensity along the radial
direction from A3558 should monotonously decrease in accordance with
the $\beta$ model profile.
Thus, we examined the unabsorbed intensity profile along the radial
direction of A3558 with 4 annular regions which are 0.54 to 0.74, 0.74
to 0.92, 0.92 to 1.1, and 1.1 to 1.24 $R_\mathrm{vir}$, respectively.
The extracted regions and the intensities are shown in Figure
\ref{fig:radial-profile-flux-comparison}.
In this analysis, we fixed the X-ray background emission using the
best-fit parameters shown in Table \ref{table:onfilament_offset-4deg} and \ref{table:onfilament_offset1} (right)
which was evaluated from the data in the entire FOV\@.  
We confirmed that resultant intensities were consistent with each other 
between the OFFSET-1deg and the OFFSET-4deg background templates in each region. 
Figure \ref{fig:radial-profile-flux-comparison} (right) shows the intensity using the OFFSET-1deg background template 
and that the intensity decrease from the first to the second bins.

For comparison, we plotted in Figure
\ref{fig:radial-profile-flux-comparison} the surface brightness of
A3558 in \cite{a3558-sb} extrapolated from the inner region data within
$\sim 0.55 R_\mathrm{vir}$ (red solid curve).  Bump-like structures in
this profile are caused by the ellipticity of A3558.
Clearly, the simple extrapolated intensity gives significantly 
brighter emission than the observed data.
This is possibly caused by a decrease of temperature and/or 
density in the outer region.  We modified the $\beta$ value from the
best-fit value 0.61 in \cite{a3558-sb} to 0.65 and 0.70
whereby conserving the total luminosity within $0.55 R_\mathrm{vir}$ to
be consistent with the data.
These brightness curves are also indicated in Figure
\ref{fig:radial-profile-flux-comparison}.  Thus if the observed excess
emission is originated from the outer region of A3558, either the slope
should become steeper or the intensity should drop by a factor of
$\sim 3$ below the simple extrapolated level.

The rightmost bin in Figure \ref{fig:radial-profile-flux-comparison},
which is the farthest from A3558, shows an intensity rise and requires
some additional component.  Therefore, we estimated the influence from
another nearby cluster A3556, in which its surface brightness was
derived from the index of optical galaxy distribution 
whose b$_j$ magnitude was less than 18
reported in \cite{a3556-sb} and the X-ray luminosity by
\citet{a3558-akimoto}. We assumed that the X-ray brightness is
approximated by a $\beta$ model and tried $3\ \beta$ values of 0.5,
0.6 and 0.7 by conserving the total luminosity to be the same.
\cite{fukazawa-2004} reported the $\beta$ value of A3556 to be much
smaller ($0.19^{+0.06}_{-0.02}$) than our assumption.  This result may
be caused by the faintness of this cluster.  Thus in this
analysis, we assumed typical $\beta$ values.
These $\beta$-model profiles are shown in Figure
\ref{fig:radial-profile-flux-comparison} (right) by magenta, cyan and
orange curves for $\beta = 0.5$, 0.6 and 0.7, respectively.
Observed excess emission can be expressed by a superposition of the
emission from the nearby clusters A3556 and A3558.  However, to
explain the observed data, the slope of A3558 required in the outer
region needed to be steeper than that in the inner region or the
cluster emission fell down more rapidly by a factor of $\sim 3$ at the
outer region.  Again this may be caused by a decrease of either
temperature or density in the outer region.

Next, we evaluated the possible contamination from the inner bright
emission of A3558 using {\it xissim} ftool in each bin.  As the
emission model, the thin thermal plasma model {\it apec} was employed
with the parameters set to be the best-fit values obtained by ASCA
(\cite{a3558-akimoto}, $kT = 5.6$ keV, $Z = 0.34 Z_\odot$) within
$12'$ from the center of A3558\@.  Spatial distribution of photons was
weighted by the assumed surface brightness ($\beta$-model distribution
for A3556 and an elliptical $\beta$-model distribution for A3558 as
described above), and the number of photons was equivalent to 10 Ms
exposure to keep the statistical error small.
We examined the original position of all the simulated photons and
extracted the ratio of photon counts originated from the inner cluster
region to the total detected ones in the energy range 0.5 to 2.0
keV\@.
As a result, the contribution from the inner region is $< 30\%$ (less
than $10\%$ in most of the regions in the FOV) in the observed region. 

Thirdly, we focused on the influence of unresolved point sources.
According to \cite{a3558-sb}, 22 other X-ray sources around
the A3558 cluster were detected and their detection limit was $2.9
\times 10^{-14}$ ergs s$^{-1}$ cm$^{-2}$ in 0.5-2.0 keV\@.
Among them, only 1 source with $F_X = 6.4 \times 10^{-14}$ ergs
s$^{-1}$ cm$^{-2}$ was in the FOV of {\it Suzau} ON-FILAMENT
region. This source was already excluded in our analysis.  In \cite{a3556-sb}, 
the galaxy density of A3556 and A3558 is about 0.01
and 0.05 arcmin$^{-2}$ respectively in our observed region and thus
the expected total number of galaxies in the {\it Suzaku} FOV is about
10.  Therefore, if each galaxy has the flux of the detection limit of $2.9\times10^{-14}$ ergs s$^{-1}$
cm$^{-2}$ in 0.5-2.0 keV, the total flux as the sum of these galaxies
can not reach the observed flux of $\sim3.5 \times 10 ^{-13}$ ergs
s$^{-1}$ cm$^{-2}$.
As for the forth possibility of the plasma associated with the
supercluster, although the observed excess can be expressed by a
superimposed emission from A3556 and A3558, this possibility is not
excluded.

As shown above, the observed excess emission can be expressed either
by the superposition of emission from nearby clusters A3556 and A3558
or by the supercluster plasma.  Since the supercluster space would not
be virialized yet, the observed temperature of the excess emission,
$2.3^{+0.9}_{-0.6}$ or $2.0^{+0.8}_{-0.5}$ keV in the case the OFFSET-1deg and the OFFSET-4deg background templates, 
is probably too high for the supercluster
plasma. On the other hand, this temperature is about $40\%$ of the
level measured in the central 12 arcmin region of A3558 (\cite{a3558-akimoto}). 
This ratio is consistent with the other recent
results for cluster outskirts around the virial radius
(\cite{george-pks0745-191,virial1,virial2,Simionescu-2011}). If this
excess is caused by the superposition of two nearby clusters A3556 and
A3558, we expect no temperature jump in the observed region. This is
consistent with the ON-FILAMENT results.
\subsection{Constraint on the WHIM}
\label{sec:constraint-whim-from}

\begin{figure*}
\begin{center}
\FigureFile(70mm,70mm)
  {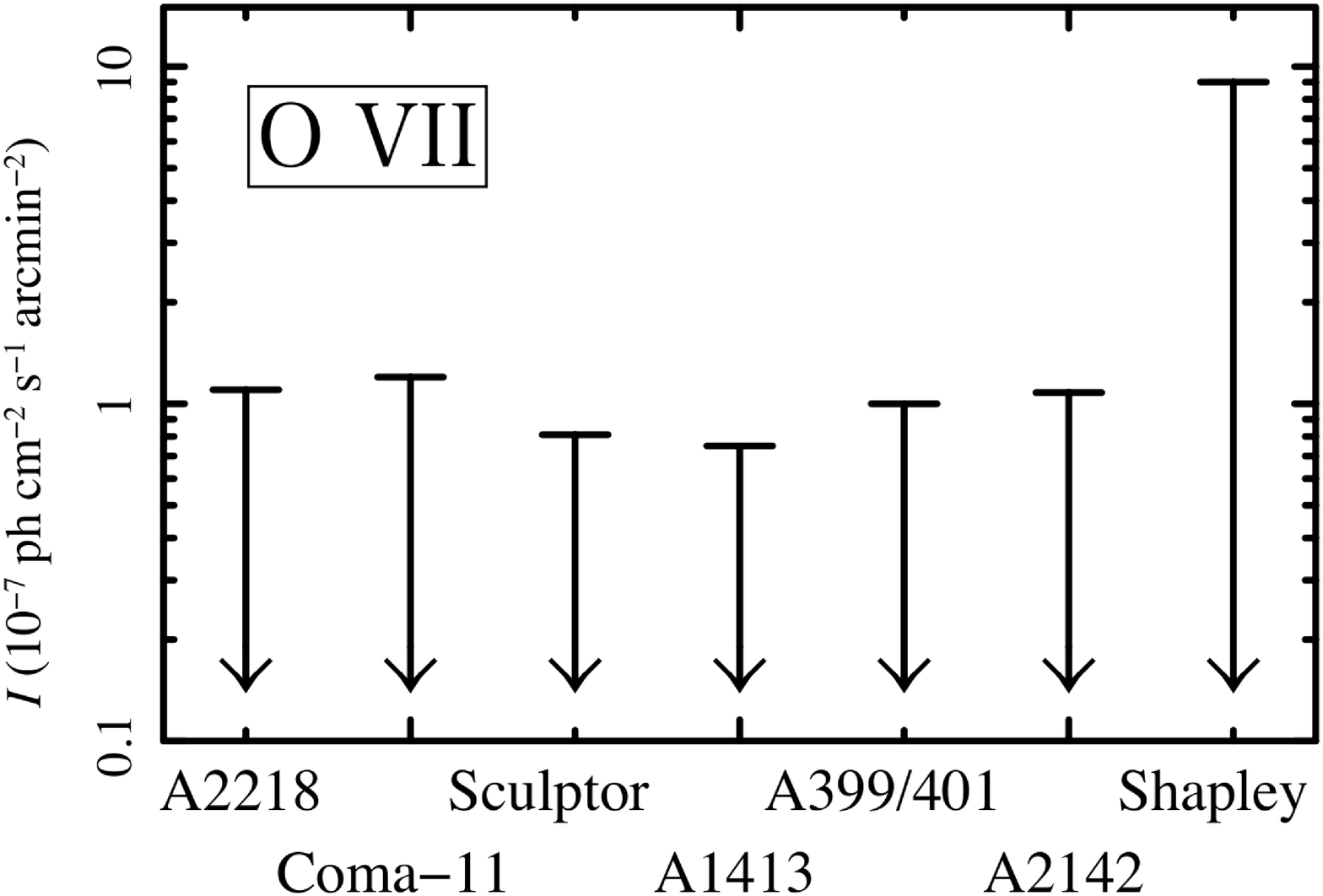}
\hspace{1cm}
\FigureFile(70mm,70mm)
  {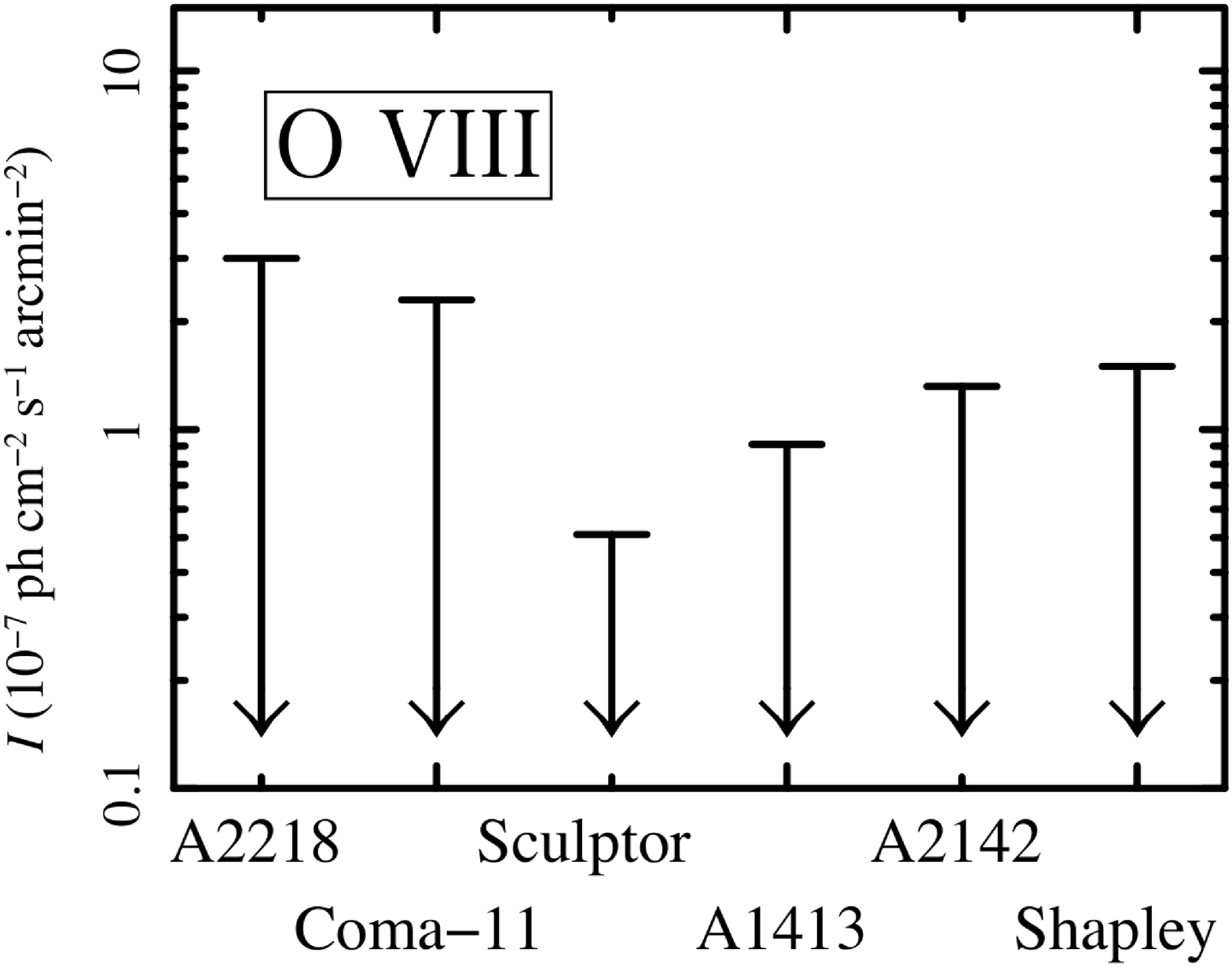}
\end{center}
 \caption{Upper limits of redshifted Oxygen emission lines from {\it Suzaku}
 observations.
Left: O\emissiontype{VII}; right: O\emissiontype{VIII}.
References are: A2218 (\cite{intensity-overdensity}),
Coma-11 (\cite{2007ApJ...655..831T}),
Sculptor (\cite{virial3}),
A1413 (\cite{virial2}), 
A399/401 (\cite{fujita-a399-a401}) and A2142 \citep{a2142-akamatsu}.
Shapley data are from this work.
}
 \label{fig:oxygen-upper-limits}
\end{figure*}

Redshifted O emission lines from the WHIM have been searched for with
{\it Suzaku} in several clusters of galaxies and superclusters.
Although the sensitivity of {\it Suzaku} is better than previous
missions, no positive detection has been obtained.
Figure~\ref{fig:oxygen-upper-limits} summarizes the upper limits of
O\emissiontype{VII} and O\emissiontype{VIII} surface brightness
obtained with {\it Suzaku}.  The data for Shapley is from ON-FILAMENT
observation of this work, and the others are from literature
(\cite{intensity-overdensity,2007ApJ...655..831T,virial3,virial2,fujita-a399-a401,a2142-akamatsu}).
The sensitivity depends on observation dates due to degradation and
uncertainty of effective area in the low energy band.  It also depends
on the redshift of objects, because Galactic strong
O\emissiontype{VII} line may overlap with the redshifted
O\emissiontype{VIII} line for some redshift range.  The latter is the
reason of relatively high upper limits of the O\emissiontype{VII} line
in the Shapley supercluster.  If we exclude this unfortunate case, the
upper limits in general reach $\sim 10^{-7}~\mathrm{photons~ cm^{-2}~
  s^{-1}~ arcmin^{-2}}$ for both O\emissiontype{VII} and
O\emissiontype{VIII} lines.  The fact that redshifted O lines have not
so far been significantly detected indicates that the average
temperature and density of the WHIM associated with clusters of galaxies or
superclusters lie outside of the detectable range.

\begin{figure*}
\begin{center}
\FigureFile(75mm,75mm)
  {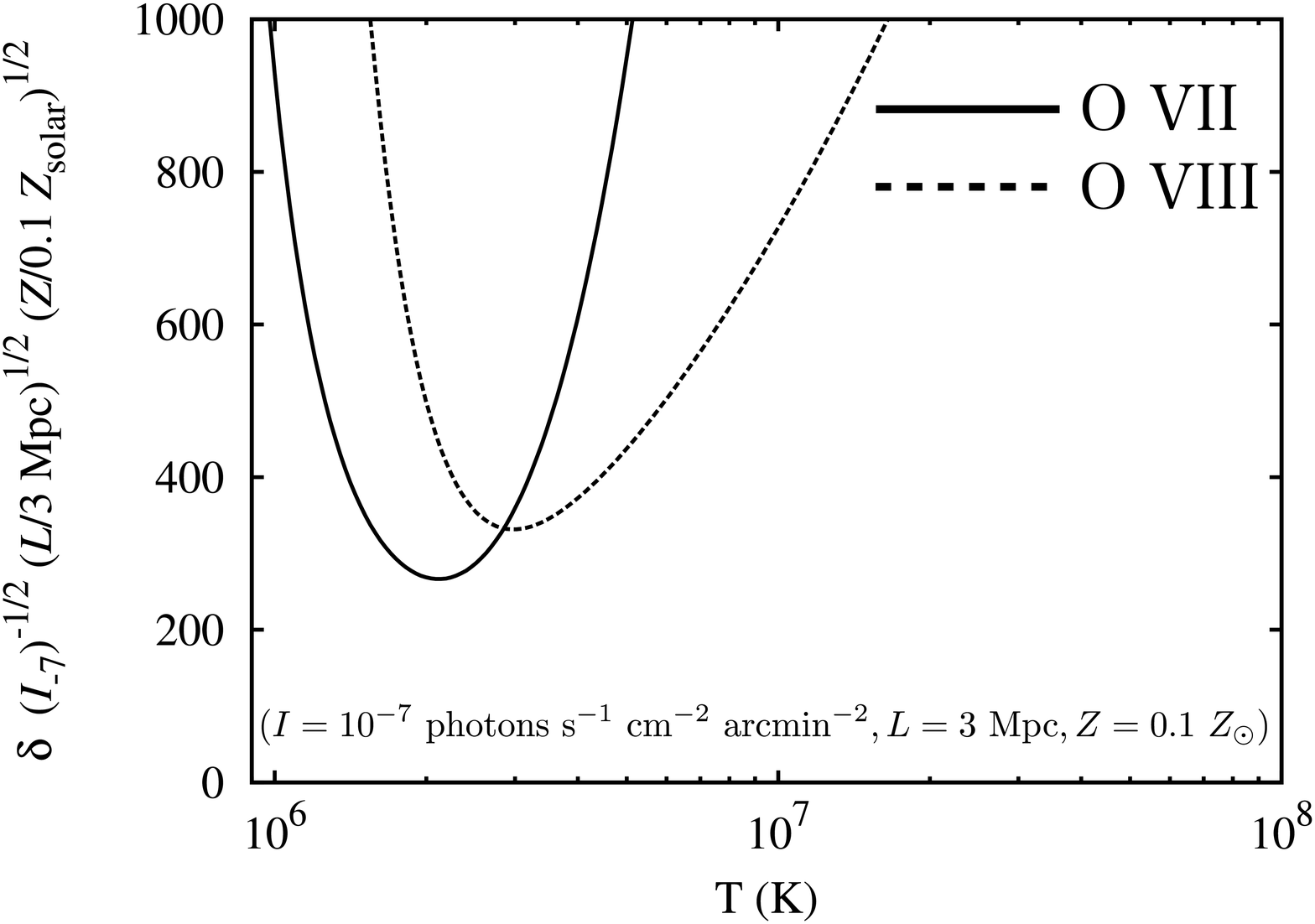}
\hspace{0.03\textwidth}
\FigureFile(75mm,75mm)
  {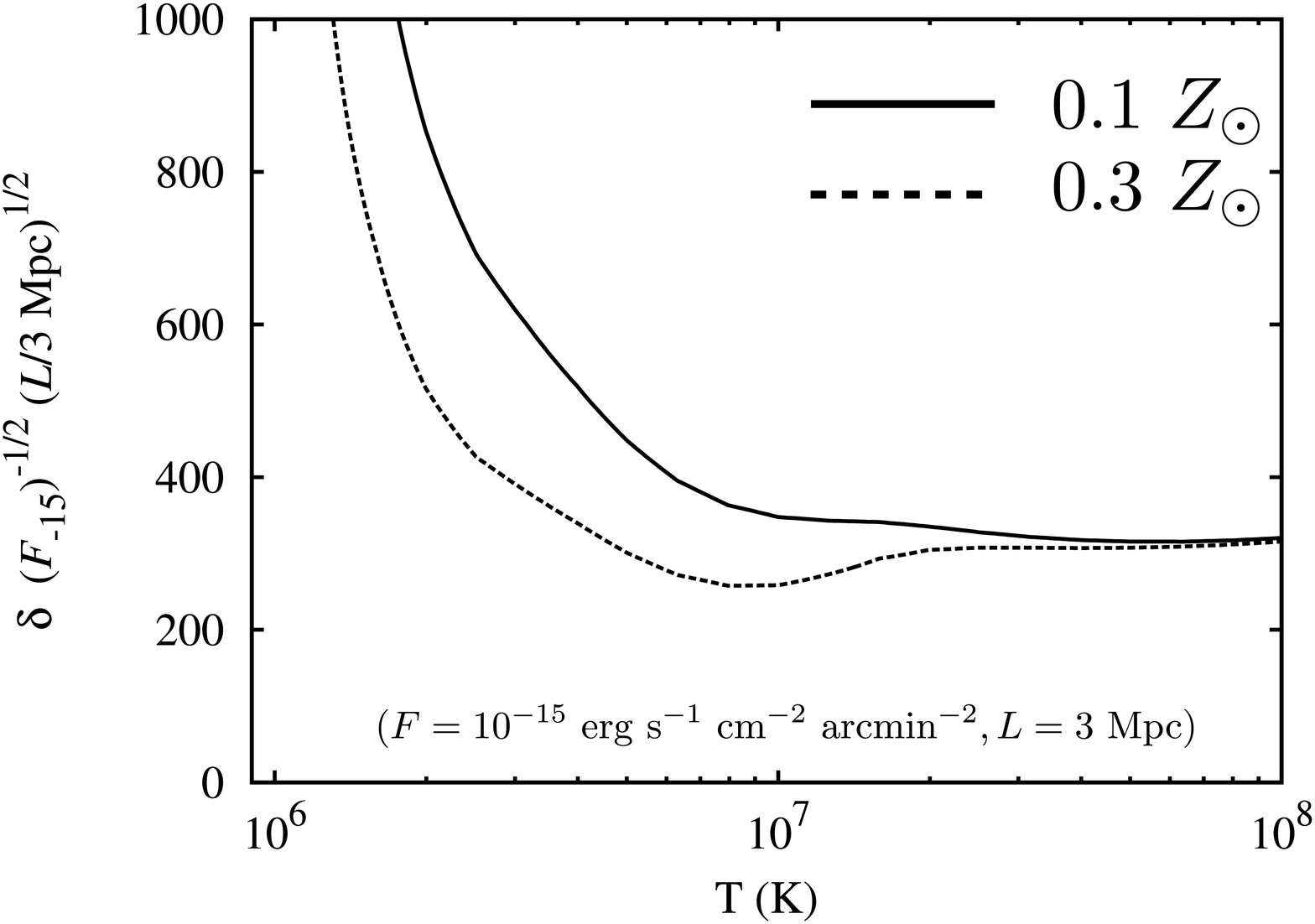}
\end{center}
\caption{ The upper limit of overdensity ($\delta$) as a function of
  temperature.  Left: the upper limit from O\emissiontype{VII} (solid)
  and O\emissiontype{VIII} (dashed) emission lines.
  $I=10^{-7}$~photons s$^{-1}$ cm$^{-2}$ arcmin$^{-2}$ ($I_{-7}=1$),
  $L=3$~Mpc and $Z=0.1~Z_\mathrm{\odot}$ are assumed.  Right: the
  upper limit from continuum emission.  $F=10^{-15}$~ergs s$^{-1}$
  cm$^{-2}$ arcmin$^{-2}=1$ ($F_{-15}=1$) and $L=3$~Mpc are assumed.
  Solid and dashed lines are with $Z=0.1~Z_{\odot}$ and
  $Z=0.3~Z_\odot$, respectively.  }
 \label{fig:whim-phase-upper}
\end{figure*}

In the calculation of the upper limit of $\delta$ shown in
\S~\ref{SEC:upper-limit-intens}, we assumed a fixed temperature of
$T=3\times10^6$~K\@.  The left panel of
Figure~\ref{fig:whim-phase-upper} shows our calculation showing the
temperature dependence under the assumption of CIE, based on the SPEX
code (\cite{spex}).  The plot shows $\delta~ ({I_{-7}})^{-1/2}~
(L/3~\mathrm{Mpc})^{1/2}~ (Z/0.1 Z_\odot)^{1/2}$ as a function of
temperature, where $I_{-7}= I/10^{-7}~ \mathrm{photons~ cm^{-2}~
  s^{-1}~ arcmin^{-2}}$.  Since $\delta$ is proportional to $I^{1/2}~
L^{-1/2}~ Z^{-1/2}$ for a fixed temperature, the curves in the plot
correspond to the upper limit of $\delta$ assuming $I_{-7}=1$,
$L=3$~Mpc, and $Z=0.1~Z_\odot$.  The solid line indicates the upper
limit determined by the O\emissiontype{VII} line, while the dashed
line is by the O\emissiontype{VIII} line.  When both the
O\emissiontype{VII} and O\emissiontype{VIII} lines give the same upper
limits, the O\emissiontype{VII} constraint is tighter in the
temperature range $T<3\times10^{6}$~K, while O\emissiontype{VIII} is
more sensitive in the higher temperatures.  The {\it Suzaku}
detectability ($I_{-7}\lesssim1$ for both O\emissiontype{VII} and
O\emissiontype{VIII}) corresponds to $\delta<400$ in a relatively
wide temperature range: $1.4\times 10^6~\mathrm{K}<T<5\times
10^{6}~\mathrm{K}$.

In the higher temperature range where the O lines are not sensitive
(i.e., $T>5\times10^6$~K), we expect continuum emission to give
a better constraint.  The right panel of
Figure~\ref{fig:whim-phase-upper} shows $\delta~ ({F_{-15}})^{-1/2}~
(L/3~\mathrm{Mpc})^{1/2}$ as a function of temperature, where
$F_{-15}= F/10^{-15}~ \mathrm{ergs~ cm^{-2}~ s^{-1}~ arcmin^{-2}}$ and
$F$ is the flux in 0.5--2.0~keV\@.  Solid and dashed lines are for
$Z=0.1~Z_\odot$ and $Z=0.3~Z_\odot$, respectively.

The flux of the continuum observed in the ON-FILAMENT region is
$F_{-15}=1.0\pm 0.2$.  Hence, the continuum from $\delta\lesssim 400$
and $T\gtrsim 5\times10^6$~K plasma can lie statistically above the
detection limit of {\it Suzaku}.  However, as discussed in
\S~\ref{SEC:orig-therm}, the origin of such a weak continuum is hard
to be identified, because of the uncertainty in the Galactic emission
and of the difficulty in spectroscopically constraining the redshift
with the CCD energy resolution.  Although the detection limit
considering the systematics is not clear, there have been no strong
indications of $T<10^7$~K continuum associated with clusters or
superclusters from {\it Suzaku} observations.

To summarize the search for the WHIM signal with {\it Suzaku}, we can
exclude the temperature-density region with $1.4\times10^6~\mathrm{K}
<T<5\times10^6$~K and average $\delta>400$ or with $T<10^7$~K and
average $\delta\gg400$. Such WHIM gas is not typically associated with
clusters of galaxies and superclusters.

\citet{werner} detected continuum emission in a bridge structure
between A222 and A223\@.  The spectrum can be represented by a CIE
plasma with $kT=0.9$~keV and $\delta=150$, assuming $L=15$~Mpc.
Because of the very large line-of-sight depth, the density is lower
than the detection limit shown in Figure~\ref{fig:whim-phase-upper}
and the temperature is higher than the level effectively probed with O
emission lines.  This result also suggests that the overdensity of the
intergalactic plasma, especially with $T\sim$(2--3)$\times10^6$~K, is
less than 400, which is consistent with our {\it Suzaku} observations.

There are several other claims from CCD observations of possible
detection of WHIM signals including O lines (e.g.,
\cite{kaastra-whim,finoguenov-whim}).  The reported
intensities of O lines are stronger than the upper limit from {\it
  Suzaku}.  The reason of this discrepancy is thought to be the
overestimation of intergalactic emission in those reports, due to
insufficient modeling of the detector response and background, or due
to spatial and temporal variation of foreground emission (e.g.,
\cite{2006ApJ...644..167B,2008ApJ...680.1049T}).  Hence, the true
signals from the WHIM are likely to be lower than the upper limits
obtained with {\it Suzaku}.

The observational constraint on $\delta$ can be compared with
predictions from cosmological numerical simulations.  Recently many
extensive simulation studies have been performed by several groups
with various physical processes considered as controlling parameters
(e.g., 
\cite{2010MNRAS.407..544B,2010MNRAS.402.1536S,2010MNRAS.409..132W,2010MNRAS.402.1911T,ursino2,2011MNRAS.415..353W}).
Although the phase diagram of the WHIM depends on the details of
physical processes such as galactic wind, AGN feedback, metal line
cooling, photoionization by background radiation, and sub-resolution
turbulence, the simulations in general predict the high-density end of
the WHIM to have $\delta>400$ and $T>10^6$~K.  However, the simulations
also predict that such high density regions are typically smaller than
the integrated volume with current {\it Suzaku} observations.  This is
qualitatively seen by comparing Figure~6 of \citet{2009ApJ...697..328B}
and Figure~4 (top left) of \citet{2011ApJ...734...91T}.  Both plots
show the phase diagram of the WHIM from the same simulation, but the
latter shows average quantities in a 500~kpc $\times$ 500~kpc $\times$
3~Mpc volume, which is similar to the integrated volume in this work
($\sim1$~Mpc $\times$ 1~Mpc $\times$ 3~Mpc).  Aa a result of the
averaging, no WHIM can be identified to have $\delta>400$ and
$T<10^7$~K\@.  We should emphasize that the plot is just the result from
one particular simulation, and different simulation parameters (i.e.,
different assumption of the physical processes) may lead to different
results.  Nevertheless, we expect this conclusion to hold for
a wide range of input parameters, at least in a qualitative manner.
Simulation in \citet{ursino1} also indicates that the continuum
emission from the WHIM, after integrating $>10'\times10'$ FOV, is smaller than
$20~\mathrm{photons~ s^{-1}~ cm^{-2}~ sr^{-1}}$ (0.375--0.950~keV), 
which is consistent with this work.

The discussion indicates that the clear detection of the WHIM X-ray
signals is challenging with CCD cameras.  Spatially resolved
spectroscopy for $L\gtrsim10$~Mpc path length may work as shown in
\citet{werner}, but this requires a fortunate orientation of the
cosmological filamentary structure.  Otherwise, the sensitivity on
the surface brightness needs to be improved by $\sim$one order of
magnitude, considering $I\propto \delta^2$.  This will be achieved by
future missions equipped with a microcalorimeter array, which enables
us to pick up weak redshifted O lines by an excellent energy
resolution of $\Delta E =$ a few eV\@.  Dedicated X-ray missions have
been proposed with instruments with a few eV energy resolution and
$\sim$deg$^2$ FOV, such as {\it DIOS}, {\it EDGE}, {\it XENIA} and
{\it ORIGIN}
(\cite{2010SPIE.7732E..54O,2009ExA....23...67P,2010SPIE.7732E..55B,denherder-2011}). 
Even though these missions are not formally approved for future space
program, we stress that such wide-field missions are capable of not
only detecting the WHIM but also studying its 3D structure by mapping
the redshifted O emission lines (\cite{2011ApJ...734...91T}).

\section{Conclusions}
We examined the X-ray spectral properties of inter-cluster regions in
the Shapley supercluster based on three {\it Suzaku} observations
named as the OFFSET-4deg, OFFSET-1deg and ON-FILAMENT regions.
Although redshifted Oxygen emission lines were not detected, the
OFFSET-1deg and the ON-FILAMENT regions indicated excess continuum
emission compared with the OFFSET-4deg region whose spectrum could be
fitted with the typical model for blank skies. These additional
emissions were represented by thermal models with $kT\sim 1$ keV and
$\sim 2$ keV in the OFFSET-1deg and the ON-FILAMENT regions,
respectively.  For the origin of the 1 keV emission, Galactic (Ne-rich
or high temperature) and supercluster plasma were both statistically
acceptable.  The 2 keV plasma between A3556 and A3558 is likely to be
the ICM or supercluster origin considering the radial intensity
profile.  However, the obtained temperature of $2.3^{+0.9}_{-0.6}$ and $2.0^{+0.8}_{-0.5}$ keV 
is too high for a supercluster plasma and the observed drop of
temperature is consistent with the {\it Suzaku} measurements of other
clusters.  Assuming this plasma to be the WHIM bound in the region
between A3556 and A3558, we derived an upper limit for the
redshifted O\emissiontype{VIII} intensity to be $1.5 \times 10^{-7}$ photons
s$^{-1}$ cm$^{-2}$ arcmin$^{-2}$. This corresponds to an overdensity
of $\sim 380 (Z/0.1~Z_{\odot})^{-1/2}\ (L/3~{\rm{Mpc}})^{-1/2}$,
assuming $T=3\times10^6 $K\@.

Finally, we summarized the previous {\it Suzaku} observations on the
WHIM search including this work and discussed the feasibility about
the constraint on the overdensity of the WHIM using {\it Suzaku}
XIS\@. Consequently, an overdensity of $< 400$ can be detectable using
O\emissiontype{VII} and O\emissiontype{VIII} emission lines for a
temperature range $1.4\times 10^6~\mathrm{K}<T<5\times
10^{6}~\mathrm{K}$ and a continuum emission fof $T > 5\times 10^6$
K\@. Considering the {\it Suzaku} results so far obtained, an
interpretation that typical line-of-sight integrated average
overdensity is $< 400$ can be derived. 

\section*{Acknowledgement}
I. M. is grateful to Kentaro Someya, Hiroshi Yoshitake, Kazuhiro Sakai and Prof. Kazuhisa Mitsuda for useful advice and discussion. 
Part of this work was financially supported by the Ministry of Education, Culture, Sports, Science and 
Technology, Grant-in Aid for Scientific Research 
10J07487, 15340088, 20340041, 21224003 and 22111513.



\end{document}